\documentclass[journal=jctcce,amsmath,amssymb,manuscript=article]{achemso}
\usepackage{amsmath}
\usepackage{subcaption}
\usepackage{graphicx}
\usepackage{subcaption}
\usepackage{multirow}
\usepackage{braket}
\usepackage{color}
\usepackage[normalem]{ulem}
\usepackage{booktabs,siunitx,array}
\usepackage{booktabs}
\usepackage{ulem}
\usepackage{listings}
\usepackage{import}
\usepackage{xcolor}
\usepackage{framed}
\usepackage{mdwlist}
\usepackage{diagbox}
\usepackage{bm}
\usepackage{graphicx}
\usepackage{dcolumn}
\usepackage{algorithm}
\usepackage{algorithmicx}
\usepackage{algpseudocode}
\usepackage{csquotes}
\usepackage{array}
\usepackage{siunitx}
\usepackage{amsmath}

\usepackage{listings}
\usepackage{xcolor}

\title{Efficient hybrid-functional-based force and stress calculations for periodic systems with
thousands of atoms}

\author{Peize Lin}
\affiliation{Institute of Physics, Chinese Academy of Sciences, Beijing 100190, China}
\alsoaffiliation{Songshan Lake Materials Laboratory, Dongguan 523808, Guangdong, China}
\alsoaffiliation{Institute of Artificial Intelligence, Hefei Comprehensive National Science Center, Hefei 230026, Anhui, China.}

\author{Yuyang Ji}
\affiliation{CAS Key Laboratory of Quantum Information, University of Science and Technology of
China, Hefei 230026, Anhui, China}

\author{Lixin He}
\affiliation{CAS Key Laboratory of Quantum Information, University of Science and Technology of
China, Hefei 230026, Anhui, China}
\alsoaffiliation{Institute of Artificial Intelligence, Hefei Comprehensive National Science Center, Hefei 230026, Anhui, China.}
\email{helx@ustc.edu.cn}

\author{Xinguo Ren}
\affiliation{Institute of Physics, Chinese Academy of Sciences, Beijing 100190, China}
\email{renxg@iphy.ac.cn}

\begin{document}

\newcommand{\defeq}{\ensuremath{\stackrel{\text{def}}{=}}}
\newcommand{\suml}[0]{\ensuremath{\sum\limits}}
\newcommand{\exx}[0]{\ensuremath{X}}
\newcommand{\fracp}[2]{\frac{\partial#1}{\partial#2}}
\newcommand{\br}[0]{\ensuremath{\mathbf{r}}}
\newcommand{\bF}[0]{\ensuremath{\mathbf{F}}}
\newcommand{\bfR}{ {\bf R}} \newcommand{\bfq}{ {\bf q}} \newcommand{\bfp}{ {\bf p}} \newcommand{\bfk}{ {\bf k}} \renewcommand{\matrix}[1]{\left[\begin{array}#1\end{array}\right]}

\begin{abstract}

We present an efficient linear-scaling algorithm for evaluating the analytical force and stress contributions derived from the exact-exchange energy, a key component in hybrid functional calculations. The algorithm, working equally well for molecular and periodic systems, is formulated within the framework of numerical atomic orbital (NAO) basis sets and takes advantage of the localized resolution-of-identity (LRI) technique for treating the two-electron Coulomb repulsion integrals.
The linear-scaling behavior is realized by fully exploiting the sparsity of the expansion coefficients resulting from 
the strict locality of the NAOs and the LRI ansatz. Our implementation is massively parallel, and 
enables efficient structural relaxation based on hybrid density functionals for bulk materials containing thousands of atoms.
In this work, we will present a detailed description of our algorithm and benchmark the performance of our implementation 
using illustrating examples. By optimizing the structures of the pristine and doped halide perovskite
material CsSnI$_3$ with different functionals, we find that in the presence of lattice strain, hybrid functionals provide a more accurate description of the stereochemical expression of the lone pair.

\end{abstract}

\section{Introduction}
Hybrid density functional (HDF) approach, since its inception in 1990's \cite{Becke:1993}, has prevailed in quantum chemistry for calculating molecules' properties.  The applications of HDFs to condensed matter materials \cite{Heyd/Scuseria/Ernzerhof:2003}, have shown great promise in
improving the description of both the electronic band structures and the cohesive properties \cite{Paier/etal:2006,Janesko/Henderson/Scuseria:2009}. Despite in great needs, the widespread applications of HDFs to materials in condensed phase have been severely impeded 
by their significantly increased computational cost due to the
requirement for evaluating the non-local Hartree-Fock (exact) exchange (HFX) contribution. 
The canonical scaling for evaluating the HFX potential and energy scales as $O(N^4)$. In typical situations,
the computational cost for evaluating the HFX term is two orders of magnitude or even higher than that of its local and semilocal counterparts. 

To address such computational challenges, various numerical techniques have been developed in recent years to speed up
the HFX calculations \cite{Wu/Selloni/Car:2009,Guidon/etal:08,Guidon/etal:2009,Shang/Li/Yang:2010,Qin/etal:2015,Levchenko/etal:2015,LinLin:2016,Lin/Ren/He:2020,Lin/Ren/He:2021}, reducing the canonical $O(N^4)$ scaling to asymptotically linear scaling.  In particular, using the localized variant of the resolution of identity (RI) approach \cite{Billingsley/Bloor:1971,Pisani/etal:2005,Pisani/etal:2008,Sodt/etal:2006,Sodt/Head-Gordon:2008,Reine/etal:2008,Merlot/etal:2013,Ihrig/etal:2015}, the amount of Coulomb integrals that need to be computed is drastically reduced. Exploiting the sparsity of the RI expansion coefficients and the
density matrix, one can design efficient linear-scaling algorithms for evaluating the HFX with rather small prefactors, enabling routine hybrid functional calculations for periodic systems containing hundreds to thousands \cite{Levchenko/etal:2015,Lin/Ren/He:2021,Lee/Head-Gordon/etal:2022,Bussy/Hutter:2024} or even ten thousands of atoms \cite{Kokott/etal:2024}. It should be noted that the speed-up gained in the HDF calculations does not come with sacrificing  
the numerical accuracy. 
According to a recent benchmark study for semiconductor materials across four independent implementation of the screened Heyd-Scuseria-Ernzerhof 
(HSE) \cite{Heyd/Scuseria/Ernzerhof:2003} hybrid functionals, the major differences in the calculated cohesive
properties stems from the pseudopotentials describing the core-valence interactions, rather than the type of basis sets and/or 
the employed localized RI (LRI) approximation.
In analogy to the algorithmic development within the atom-centered atomic-orbital (AO) framework, 
improved low-scaling algorithm has also been developed for
the plane-wave basis sets, utilizing the interpolative separable density fitting technique \cite{Lu/Yin:2015,Carnimeo_2019,Hu/Lin/Yang:2017,Wu/Qin/etal:2022}.

Again, due to the significantly increased computational cost associated with HFX, relaxing the atomic geometries using
HDFs is even more computationally challenging. Thus, an often chosen protocol in
computational materials science is to first relax the structures using semi-local functionals, and subsequently 
calculate the electronic band structures using more advanced HDFs at fixed geometry. This approach works rather well for
a large variety of materials. However, there exists situations where the electronic structure is
rather sensitive to the details of the atomic positions; as a result, calculations based on the GGA geometry or HDF geometry
can lead to substantial difference in the band gap. Certain halide double perovskite 
photovoltaic materials \cite{Ji/etal:2024} and polarons in ionic materials \cite{Franchini2021}
are good examples of this kind. 
In such situations, it is highly
desirable to be able to efficiently relax the geometrical structures consistently using HDFs.

In this work, building on top of our previous efficient HFX implementation within the framework of numerical atomic orbital (NAO) basis set,
we extend our formalism and implementation to the analytical force and stress calculations associated with the HFX term. This
extension allows us to relax the geometrical structures of condensed materials using HDFs rather efficiently. In the following, the paper is organized as follows. In Sec.~\ref{sec:methods} the formulation and key equations of the HFX forces and stresses under the LRI ansatz within the NAO basis-set framework are presented. The algorithms behind our implementation, in particular regarding the detailed procedures to exploit the sparsity of the intermediate quantities and the hybrid MPI/OpenMP parallelization schemes, are discussed in Sec.~\ref{sec:algorithm}. Sec.~\ref{sec:results} presents the results of this work, including the validity check of our analytical force and stress implementation against the finite difference results, the efficacy
of our screening algorithm and the scaling behavior of our implementation with respect to 
system size and CPU cores. An application of our code to study the lone-pair stereochemical effect
in halide perovskite material is also reported in Sec.~\ref{sec:results}.
Finally, we conclude this work in Sec.~\ref{sec:summary}.

\section{\label{sec:methods} Methods}

Our approach is formulated within the framework of AO basis sets, whereby
the sparsity arising from the locality of AOs can be exploited.
We denote a localized atomic basis function, centered on an atom $\tilde{I}$ inside a unit cell specified by
the Bravais lattice vector $\mathbf{R}$, as
\begin{equation}
	\phi_{i,\tilde{I}(\mathbf{R})}(\br) \defeq \phi_i(\br-\bm{\tau}_{\tilde{I}}-\mathbf{R}) \, ,
\end{equation}
where $\bm{\tau}_{\tilde{I}}$ is the position of the atom $\tilde{I}$ within the unit cell. 
To simplify the notation, we follow Ref.~\citenum{Lin/Ren/He:2021}
by combining the lattice vector $\mathbf{R}$ and atom $\tilde{I}$ into a single index $I$ in subsequent discussions, 
\begin{equation}
	\phi_{Ii}(\br) \defeq  \phi_{i,\tilde{I}(\mathbf{R})}(\br).
\end{equation}

In terms of the AO basis sets $\{\phi_{Ii}(\br)\}$,
the HFX energy per unit cell can be obtained as
\begin{equation}
	E^\exx = -\frac{1}{2N_k} \suml_{Ii,Jj,Kk,Ll} (\phi_{Ii}\phi_{Kk}|\phi_{Jj}\phi_{Ll}) D_{Kk,Ll} D^*_{Ii,Jj}
\label{Eexx} ,\end{equation}
where $N_k$ is the number of $\bfk$-points in the first Brillouin zone, equivalent to the number of unit cells in a Born-von K\'{a}rm\'{a}n supercell.
Note that, in Eq.~\ref{Eexx}, the summations over atoms $I,J,K,L$ go over the BvK supercell.
$D_{Ii,Jj}$ in  Eq.~\ref{Eexx} is the density matrix
\begin{equation}
    D_{Ii,Jj}
    \defeq D_{\tilde{I}(\bfR_I)i, \tilde{J}(\bfR_J)j}
    = \frac{1}{N_k} \suml_\bfk e^{-i\bfk \cdot(\bfR_J-\bfR_I)} \suml_n f_{n\bfk} c_{\tilde{I}i,n\bfk} c_{\tilde{J}j,n\bfk}^*
\end{equation}
where $c_{\tilde{I}i,n\bfk}$ are the KS eigenvector and $f_{n\bfk}$ the occupation numbers, and the summation over $\bfk$ goes over the
first Brillouin zone. Furthermore, in Eq.~\ref{Eexx} we follow the usual convention to define the electron repulsion integrals (ERIs)  between two functions as
\begin{equation}
	(f|g) \defeq \iint f(\br) v(\br-\br') g(\br') d\br d\br' \, ,
 \label{eq:Coulomb_def}
\end{equation}
where the Coulomb potential $v(\br)$ can
be either full-ranged [i.e., $v(\br)=1/|\br|$] for Hartree-Fock
or short-ranged [e.g., $v(\br)=\text{erfc}(\omega|\br|)/|\br|$ with $\omega$ being the screening parameter] for HSE-type
screened hybrid functionals. 

The number of ERIs is proportional to the fourth power of the number of atoms.
Although the values of these ERIs depend only on the AOs and atomic positions,
and remain unchanged during the self-consistent calculations, 
the enormous number of ERIs renders a \textit{direct} approach, i.e., keeping all the ERIs in memory at the same time, unfeasible.
Furthermore, calculating the ERIs for NAOs requires a 6-dimensional integration over the grid points $\br$ and $\br'$, which is computationally extremely challenging.
To reduce the computational burden, the LRI approximation \cite{Ihrig/etal:2015, levchenko2015hybrid, Lin/Ren/He:2020} is invoked in the present work to evaluate the ERIs. The core idea of the LRI, also known as pair-atomic RI (PARI), 
pair-atomic density fitting \cite{Merlot/etal:2013,Wirz/etal:2017}, or concentric atomic density fitting \cite{Hollman/Schaefer/Valeev:2014,Wang/Lewis/Valeev:2020} in the literature, is to expand the product of two AOs 
in terms of a subset of ABFs
which are also centering on the two atoms in question,
\begin{equation}\begin{array}{lcl}
	\phi_{Ii}(\br) \phi_{Kk}(\br)
	&\approx& \suml_{A=\{I,K\}} \suml_{\alpha\in A} C_{Ii,Kk}^{A\alpha} P_{A\alpha}(\br)		\\
	&=& \suml_{\alpha\in I} C_{Ii,Kk}^{I\alpha} P_{I\alpha}(\br) + \suml_{\alpha\in K} C_{Ii,Kk}^{K\alpha} P_{K\alpha}(\br)
\end{array}\end{equation}
where $\{P_{A\alpha}(\br)\}$ are the atom-centered ABFs,
\begin{equation}
	P_{A\alpha}(\br) \defeq P_{\alpha}(\br-\bm{\tau}_{\tilde{A}}) \, .
\end{equation}
Without losing generality, the NAOs and ABFs are assumed to be real in this article. 
The expansion coefficients $C_{Ii,Kk}^{A\alpha}$ with $A=\{I,K\}$ are a two-center, three-index tensor
and apparently $C_{Ii,Kk}^{A\alpha} = C_{Kk,Ii}^{A\alpha}$.

Within the LRI, the HFX energy $E^\exx$ can be expressed as
\begin{equation}
	E^\exx = -\frac{1}{2N_\bfk} \suml_{IJKL} \suml_{A=\{I,K\}} \suml_{B=\{J,L\}} C_{Ii,Kk}^{A\alpha} (P_{A\alpha}|P_{B\beta}) C_{Jj,Ll}^{B\beta} D_{Kk,Ll} D_{Ii,Jj}^* \, ,
\label{Exx-LRI} \end{equation}
where $(P_{A\alpha}|P_{B\beta})$ is the Coulomb integrals between the ABFs, defined according to Eq.~\ref{eq:Coulomb_def}.
Below, we denote the Coulomb integral between two ABFs as
\begin{equation}
	V_{A\alpha,B\beta} \defeq (P_{A\alpha}|P_{B\beta})\, .
\label{V}\end{equation}
For brevity, in the following the tensor $C_{Ii,Kk}^{A\alpha}$ is abbreviated as $C_{IK}^{A}$, and
similarly $V_{A\alpha,B\beta}$ abbreviated as $V_{AB}$ and $D_{Kk,Ll}$ as $D_{KL}$, unless the AO indices need to be explicitly specified.

\begin{figure}[!htbp]
	\includegraphics[width=0.7\textwidth]{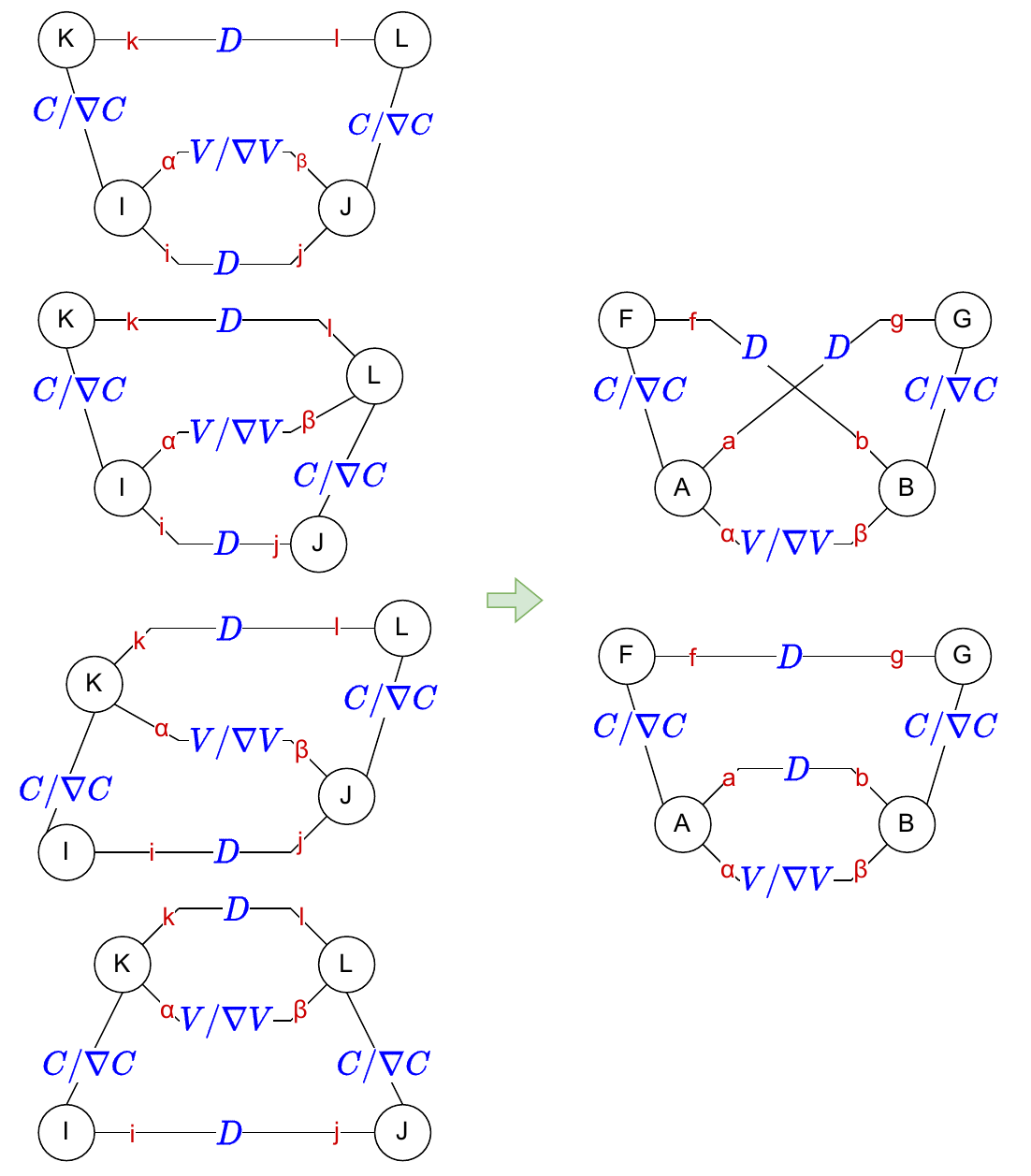}
	\centering
    \caption{Change of the perspective for building the HFX matrix. The left panel is a schematic illustration of the contraction process defined by Eq.~\ref{eq:IJKL_perspective} while the right panel is a illustration of Eq.~\ref{eq:ABFG_perspective}.
    The circles refer to atoms, and the lines refer to the tensors. The indices of NAOs and ABFs are all contracted.}
    \label{fig:change_of_perspective}
\end{figure}

Now the force acting on an atom $M$ due to the HFX energy is given by the negative of the derivative of Eq.~\ref{Exx-LRI} with respect to the nuclear coordinates, i.e., $\bF^\exx_M = -\nabla_M E^\exx =  -\nabla_{\bm{\tau}_M+\mathbf{R}_M} E^\exx$, where $\bm{\tau}_M+\mathbf{R}_M$ is the position of the atom $M$.
To obtain the concrete mathematical formula of the force and the stress, we re-express Eq.~\ref{Exx-LRI}  as
\begin{eqnarray}
	E^\exx
	&=& -\frac{1}{2} \suml_{IJKL} \suml_{A=\{I,K\}} \suml_{B=\{J,L\}} \suml_{ijkl}\suml_{\alpha\beta}
		C_{Ii,Kk}^{A\alpha} V_{A\alpha,B\beta} C_{Jj,Ll}^B D_{Kk,Ll} D_{Ii,Jj}^*	 \nonumber \\
	&=& -\frac{1}{2} \left[
		        \suml_{IJKL} \suml_{ijkl}\suml_{\alpha\beta} C_{Ii,Kk}^{I\alpha} V_{I\alpha,J\beta} C_{Jj,Ll}^{J\beta} D_{K\alpha,L\beta} D_{Ii,Jj}^* 	\right. \nonumber	\\
	&&+         \suml_{IJKL} \suml_{ijkl}\suml_{\alpha\beta} C_{Ii,Kk}^{I\alpha} V_{I\alpha,L\beta} C_{Jj,Ll}^{L\beta} D_{K\alpha,L\beta} D_{Ii,Jj}^* 	 \nonumber	\\
	&&+         \suml_{IJKL} \suml_{ijkl}\suml_{\alpha\beta} C_{Ii,Kk}^{K\alpha} V_{K\alpha,J\beta} C_{Jj,Ll}^{J\beta} D_{K\alpha,L\beta} D_{Ii,Jj}^* 	 \nonumber	\\
	&&+  \left. \suml_{IJKL} \suml_{ijkl}\suml_{\alpha\beta} C_{Ii,Kk}^{K\alpha} V_{K\alpha,L\beta} C_{Jj,Ll}^{L\beta} D_{K\alpha,L\beta} D_{Ii,Jj}^* \right]	\label{eq:IJKL_perspective}	\\
	&=& -\frac{1}{2} \left[
		        \suml_{ABFG} \suml_{abfg}\suml_{\alpha\beta} C_{Aa,Ff}^{A\alpha} V_{A\alpha,B\beta} C_{Bb,Gb}^{B\beta} D_{Ff,Gg} D_{Aa,Bb}^* 	\right. \nonumber	\\
	&&+         \suml_{AGFB} \suml_{agfb}\suml_{\alpha\beta} C_{Aa,Ff}^{A\alpha} V_{A\alpha,B\beta} C_{Gg,Bb}^{B\beta} D_{Ff,Bb} D_{Aa,Gg}^* 	 \nonumber	\\
	&&+         \suml_{FBAG} \suml_{fbag}\suml_{\alpha\beta} C_{Ff,Aa}^{A\alpha} V_{A\alpha,B\beta} C_{Bb,Gg}^{B\beta} D_{Aa,Gg} D_{Ff,Bb}^* 	 \nonumber	\\
	&&+  \left. \suml_{FGAB} \suml_{fgab}\suml_{\alpha\beta} C_{Ff,Aa}^{A\alpha} V_{A\alpha,B\beta} C_{Gg,Bb}^{B\beta} D_{Aa,Bb} D_{Ff,Gg}^*  \right]	\label{eq:ABFG_perspective}	\\
	&=& -\suml_{ABFG} \suml_{abfg}\suml_{\alpha\beta} C_{Aa,Ff}^{A\alpha} V_{A\alpha,B\beta} C_{Bb,Gg}^{B\beta} \text{Re}[D_{Aa,Bb} D_{Ff,Gg}^* + D_{Aa,Gg} D_{Ff,Bb}^*], \label{eq:exx_expand}	
\end{eqnarray}
where the four terms are the characteristic of the LRI approximation \cite{Lin/Ren/He:2020,Lin/Ren/He:2021}.
In Eq.~\ref{eq:ABFG_perspective}, $A$ and $B$ denote the atoms on which the ABFs are centered, where $F$ and $G$
are neighboring atoms of $A$ and $B$, respectively. 
From Eq.~\ref{eq:IJKL_perspective} to Eq.~\ref{eq:ABFG_perspective}, we changed the perspective from the AO-centered atoms to the ABF-centered atoms,
as illustrated in Figure~\ref{fig:change_of_perspective}. As discussed in Ref.~\citenum{Lin/Ren/He:2021}, this latter perspective facilitates the design
of linear-scaling algorithm for efficient implementation.

This total force on each atom can be divided into two parts $\bF^\exx_M = \bF^{\exx1}_M + \bF^{\exx2}_M$, where
the first part $\bF^{\exx1}_M$ derives from the variation of the ERIs (or equivalently the $C$ and $V$ tensors), while the second part $\bF^{\exx2}_M$ 
is due to the variation of the density matrix.
From Eq.~\ref{eq:exx_expand}, one can obtain the energy change due to the variations of the $C$ and $V$ tensors as
\begin{equation}\begin{array}{ll}
	\delta E^{\exx1} = -\suml_{ABFG} \suml_{abfg}\suml_{\alpha\beta}
	& [
		\delta C_{Aa,Ff}^{A\alpha} V_{A\alpha,B\beta} C_{Bb,Gg}^{B\beta}
		+ C_{Aa,Ff}^A \delta V_{A\alpha,B\beta} C_{Bb,Gg}^{B\beta}
		+ C_{Aa,Ff}^A V_{A\alpha,B\beta} \delta C_{Bb,Gg}^{B\beta}]       \\
	& \text{Re}[D_{Aa,Bb} D_{Ff,Gg}^* + D_{Aa,Gg} D_{Ff,Bb}^*] \, 
 \end{array}
 \label{Eq:Exx_energy_variation}
\end{equation}
which yields the first part of the HFX force,
\begin{equation}\begin{array}{ccl}
	\bF^{\exx1}_M
	&=& -\nabla_M E^{\exx1}		\\
	&=&   \suml_{BFG} \suml_{abfg} \suml_{\alpha\beta} \left(\nabla_M C_{Ma,Ff}^{M\alpha}\right) V_{M\alpha,B\beta} C_{Bb,Gg}^{B\beta} \text{Re}[D_{Ma,Bb} D_{Ff,Gg}^* + D_{Ma,Gg} D_{Ff,Bb}^*]		\\
	&&	+ \suml_{ABG} \suml_{abfg} \suml_{\alpha\beta} \left(\nabla_M C_{Aa,Mf}^{A\alpha}\right) V_{A\alpha,B\beta} C_{Bb,Gg}^{B\beta} \text{Re}[D_{Aa,Bb} D_{Mf,Gg}^* + D_{Aa,Gg} D_{Mf,Bb}^*]		\\
	&&	+ \suml_{BFG} \suml_{abfg} \suml_{\alpha\beta} C_{Ma,Ff}^{M\alpha} \left(\nabla_M V_{M\alpha,B\beta}\right) C_{Bb,Gg}^{B\beta} \text{Re}[D_{Ma,Bb} D_{Ff,Gg}^* + D_{Ma,Gg} D_{Ff,Bb}^*]		\\
	&&	+ \suml_{AFG} \suml_{abfg} \suml_{\alpha\beta} C_{Aa,Ff}^{A\alpha} \left(\nabla_M V_{A\alpha,M\beta}\right) C_{Mb,Gg}^{M\beta} \text{Re}[D_{Aa,Mb} D_{Ff,Gg}^* + D_{Aa,Gg} D_{Ff,Mb}^*]		\\
	&&	+ \suml_{AFG} \suml_{abfg} \suml_{\alpha\beta} C_{Aa,Ff}^{A\alpha} V_{A\alpha,B\beta} \left(\nabla_M C_{Mb,Gg}^{M\beta}\right) \text{Re}[D_{Aa,Mb} D_{Ff,Gg}^* + D_{Aa,Gg} D_{Ff,Mb}^*]		\\
	&&	+ \suml_{ABF} \suml_{abfg} \suml_{\alpha\beta} C_{Aa,Ff}^{A\alpha} V_{A\alpha,B\beta} \left(\nabla_M C_{Bb,Mg}^{B\beta}\right) \text{Re}[D_{Aa,Bb} D_{Ff,Mg}^* + D_{Aa,Mg} D_{Ff,Bb}^*]	\, .	\\
\end{array}\label{Fexx1}\end{equation}
Equation~\ref{Fexx1} can be readily obtained from Eq.~\ref{Eq:Exx_energy_variation} by identifying each of the atomic indices in $\delta C_{AF}^A$, $\delta V_{AB}$, and $\delta C_{BG}^B$ with $M$, i.e., the atom on which the force is calculated.

The second part of the HFX force $\bF^{\exx2}_M$ arises from the variation of the density matrix.
From Eq.~\ref{Eexx}, the energy change induced by the variation of the density matrix is given by
\begin{equation}\begin{array}{ccl}
	\delta E^{\exx2}
	&=& -\frac{1}{2} \suml_{IJKL} \suml_{ijkl} (\phi_{Ii}\phi_{Kk}|\phi_{Jj}\phi_{Ll}) \delta D_{Kk,Ll} D_{Ii,Jj}^*        \\
	& & -\frac{1}{2} \suml_{IJKL} \suml_{ijkl} (\phi_{Ii}\phi_{Kk}|\phi_{Jj}\phi_{Ll}) D_{Kk,Ll} \delta D_{Ii,Jj}^*		\\
	&=&  \frac{1}{2} \suml_{KL} \suml_{kl} H_{Kk,Ll}^* \delta D_{Kk,Ll}
	    +\frac{1}{2} \suml_{Ii} \suml_{ij} H_{Ii,Jj} \delta D_{Ii,Jj}^*		\\
	&=&  \frac{1}{2} \suml_{KL} \suml_{kl} H_{Ll,Kk} \delta D_{Ll,Kk}^*
	    +\frac{1}{2} \suml_{Ii} \suml_{ij} H_{Ii,Jj} \delta D_{Ii,Jj}^*		\\
	&=& \suml_{IJ} \suml_{ij} H_{Ii,Jj}^{\exx} \delta D_{Ii,Jj}^* \,
\end{array}\label{Exx2}\end{equation}
where $H_{IJ}^{\exx}$ is the exact-exchange part of the Hamiltonian whose matrix elements are given by
\begin{equation}
	H^{\exx}_{Ii,Jj} = - \sum_{KL} \sum_{kl} (\phi_{Ii}\phi_{Kk}|\phi_{Jj}\phi_{Ll}) D_{Kk,Ll} \, .
\end{equation}
In order to avoid the difficulty of directly evaluating the derivative of the density matrix
with respect to the nuclear displacement, an ingenious technique \cite{Pople/etal:1979,Soler/etal:2002} for the full generalized KS Hamiltonian $H_{IJ}^\mathrm{all}$
is usually adopted in the DFT force calculations, utilizing the following relationship
\begin{equation}
	\suml_{IJ} \suml_{ij} H_{Ii,Jj}^\mathrm{all} \delta D_{Ii,Jj}^*
    = - \suml_{IJ} \suml_{ij} E_{Ii,Jj}^* \delta S_{Ii,Jj} \, ,
\label{HD_ES} \end{equation}
where $S_{IJ}$ is the overlap matrix between the AOs, and $E_{IJ}$ the 
so-called energy-weighted density matrix defined as
\begin{equation}
    E_{Ii,Jj}
    \defeq E_{\tilde{I}(\bfR_I)i, \tilde{J}(\bfR_J)j}
    = \frac{1}{N_k} \suml_\bfk e^{-i\bfk(\bfR_J-\bfR_I)} \suml_n f_{n\bfk}
    \varepsilon_{n\bfk} c_{\tilde{I}i,n\bfk} c_{\tilde{J}j,n\bfk}^* \, .
\end{equation}
Thus the force contribution arising from Eq.~\ref{Exx2} can be most conveniently included in the usual DFT force calculations,
and hence does not need to be calculated separately here.
In the following, our attention will be focused on the calculation of the first part of the force given by Eq.~\ref{Fexx1}.

In addition to atomic forces, a closely related quantity is the stress tensor, which is needed for relaxing the lattice vectors of
periodic systems. By definition, the stress tensor is given by the derivatives of the energy with respect to the strain tensor. For the HFX energy,
this amounts to computing
\begin{equation}
\sigma_{\alpha\beta}^{\exx} = - \frac{1}{\Omega N_k} \suml_{MN}  \fracp{E^{\exx}}{{\bm \tau}_{MN}^{\alpha}} {\bm \tau}_{MN}^{\beta}\, ,
\end{equation}
where $M$ and $N$ run over all atoms, ${\bm \tau}_{MN} \defeq ({\bm \tau}_M+\bfR_M) - ({\bm \tau}_N+\bfR_N)$ is the relative position of the atom $M$ and $N$ (with $\bfR_M$ and $\bfR_N$ specifying the unit cells where the atoms $M$ and $N$ are located), 
and $\alpha,\beta$ denote the three Cartesian directions. 

Similarly to the force calculations, here we only need to deal with the component of the HFX stress arising from the variations of the AO and auxiliary basis functions (i.e. the $C$ and $V$ tensors) and not with that arising from the variation of the density matrix.  Again, starting from Eq.~\ref{eq:exx_expand}, we have
\begin{equation}\begin{array}{cccl}
    \sigma^{\exx1}_{\alpha\beta}
    &=& - \frac{1}{\Omega} \suml_{MN} & \fracp{E^{\exx1}}{{\bm \tau}_{MN}^{\alpha}} {\bm \tau}_{MN}^{\beta}     \\
    &=& \frac{1}{\Omega} \suml_{ABFG} \suml_{abfg} \suml_{\alpha\beta} &
	   \left[ \fracp{C_{Aa,Ff}^{A\alpha}}{{\bm \tau}_{AF}^{\alpha}} V_{A\alpha,B\beta} C_{Bb,Gg}^{B\beta} {\bm \tau}_{AF}^{\beta}
                + C_{Aa,Ff}^{A\alpha} \fracp{V_{A\alpha,B\beta}}{{\bm \tau}_{AB}^{\alpha}} C_{Bb,Gg}^{B\beta} {\bm \tau}_{AB}^{\beta}
                + C_{Aa,Ff}^{A\alpha} V_{A\alpha,B\beta} \fracp{C_{Bb,Gg}^{B\beta}}{{\bm \tau}_{BG}^{\alpha}} {\bm \tau}_{BG}^{\beta} \right]       \\
    & & &  \text{Re}[D_{Aa,Bb} D_{Ff,Gg}^* + D_{Aa,Gg} D_{Ff,Bb}^*] \, .
\end{array}\label{stress}\end{equation}

In the next section, we will discuss how these equations are implemented.

\section{Algorithm}
  \label{sec:algorithm}

In this section, we focus on discussing the algorithms behind our HFX force and stress implementations. We will explain how the $C$ and $V$ tensors, as well as their derivatives, are calculated within the LRI formalism.  Emphases will be placed on how to make use of the symmetry and sparsity of these tensors to design an efficient linear-scaling algorithm. We will also discuss the parallelization scheme of our implementation, which is essential for achieving the goal of relaxing large-size systems using hybrid functionals.

\subsection{Preparation of $C$, $V$, $\nabla C$, $\nabla V$ tensors}

As discussed in Sec.~\ref{sec:methods}, the key intermediate quantities in the HFX force calculations under LRI are the tensors $C$, $V$
and their derivatives $\nabla C$, $\nabla V$.
These tensors have symmetry properties that one can exploit. First, according to the definition of $V_{AB}$ in Eq.~\eqref{V}, it is obvious that $V_{BA} = V_{AB}^T$.
As for the expansion coefficients $C_{IK}^{A}$, within the framework of LRI,  these are given by the requirement of minimizing the error of ERIs \cite{Ihrig/etal:2015}, leading to
the following expression,

\begin{equation}
	\matrix{{c} C_{Ii,Kk}^{I\alpha} \\ C_{Ii,Kk}^{K\alpha}}
	= \suml_{\beta} \matrix{{cc} V_{I,I} & V_{I,K} \\ V_{K,I} & V_{K,K} }^{-1}_{\alpha,\beta}
	\matrix{{c} (P_{I\beta}|\phi_{Ii}\phi_{Kk}) \\ (P_{K\beta}|\phi_{Ii}\phi_{Kk}) } \, .
\label{C_VA}\end{equation}

In Eq.~\ref{C_VA}, $(P_{I}|\phi_{I}\phi_{K})$ are the Coulomb integrals between pair products of AOs and the ABFs,
and $V_{II}$, $V_{KK}$ and $V_{IK}$ are blocks of the Coulomb matrix on the atom $I$, on the atoms $K$, and
between the atoms $I$ and $K$, respectively. Obviously, the coefficients $C_{IK}^{A}$ (with $A=I$ or $K$) satisfy the exchange symmetry $C_{IK}^{A} = C_{KI}^{A}$.

As can be seen from Eq.~\ref{C_VA}, the entire $C$ are composed of blocks associated with each pair of atoms.
For each block and a chosen set of NAOs/ABFs, the values of $V$ and $C$ tensors depend only 
on the element types of the two atoms in question and the relative position between the two atoms.
When the two atoms are shifted by the same displacement, the values of $C$ and $V$ tensors
in the corresponding block remain unchanged.
Therefore, repetitive calculations are unnecessary
for tensors with the same elements and the same relative atomic positions. In particular, a simple application of the translational symmetry to the ``on-site" elements $V_{AA}$ and $C_{AA}^A$, which stay unchanged regardless of how the atom A moves, implies $\nabla_A V_{AA}=0$ and $\nabla_A C_{AA}^{A}=0$.
Furthermore, according to the definition $\nabla_B V_{AB} = (P_{A}|\nabla_B P_{B})$,
we have $\nabla_B V_{AB} = - \nabla_{A} V_{AB}$

In order to calculate $\nabla_K C_{IK}^A$, one first needs to determine $\nabla_B V_{AB}^{-1}$, as can be seen from
Eq.~\ref{C_VA}.
To this end, we use the relationship $V_{AB} * V_{AB}^{-1} =I$ with $I$ being the identity matrix. 
By applying the gradient operator to both sides of the identity equation, one obtains
\begin{equation}
	\nabla_B V_{AB}^{-1} = - V_{AB}^{-1} * \nabla_B V_{AB} * V_{AB}^{-1}
 \label{eq:V_inverse}
\end{equation}
	
Combining Eqs.~\ref{C_VA} and ~\ref{eq:V_inverse}, we arrive at the following expression for the gradients of the $C$ tensor,
\begin{equation}\begin{array}{ccl}
	&& \matrix{{c} \nabla_K C_{Ii,Kk}^{I\alpha} \\ \nabla_K C_{Ii,Kk}^{K\alpha}}       \\
	&=& \suml_{\beta} \nabla_K \matrix{{cc} V_{I,I} & V_{I,K} \\ V_{K,I} & V_{K,K} }_{\alpha,\beta}^{-1} * \matrix{{c} (P_{I\beta}|\phi_{Ii}\phi_{Kk}) \\ (P_{K\beta}|\phi_{Ii}\phi_{Kk}) }		\\
	&&	+ \suml_{\beta} \matrix{{cc} V_{I,I} & V_{I,K} \\ V_{K,I} & V_{K,K} }_{\alpha,\beta}^{-1} * \matrix{{c} \nabla_K (P_{I\beta}|\phi_{Ii}\phi_{Kk}) \\ \nabla_K (P_{K\beta}|\phi_{Ii}\phi_{Kk}) }		\\
	&=& - \suml_{\beta\gamma\delta} \matrix{{cc} V_{I,I} & V_{I,K} \\ V_{K,I} & V_{K,K} }_{\alpha,\beta}^{-1}
		* \matrix{{cc} 0 & \nabla_K V_{I\beta,K\gamma} \\ \nabla_K V_{K\beta,I\gamma} & 0 }
		* \matrix{{cc} V_{I,I} & V_{I,K} \\ V_{K,I} & V_{K,K} }_{\gamma,\delta}^{-1}
		* \matrix{{c} (P_{I\delta}|\phi_{Ii}\phi_{Kk}) \\ (P_{K\delta}|\phi_{Ii}\phi_{Kk}) }		\\
	&&	+ \suml_{\beta} \matrix{{cc} V_{I,I} & V_{I,K} \\ V_{K,I} & V_{K,K} }_{\alpha,\beta}^{-1} * \matrix{{c} \nabla_K (P_{I\beta}|\phi_{Ii}\phi_{Kk}) \\ \nabla_K (P_{K\beta}|\phi_{Ii}\phi_{Kk}) }		\\
	&=& \suml_{\beta} \matrix{{cc} V_{I,I} & V_{I,K} \\ V_{K,I} & V_{K,K} }_{\alpha,\beta}^{-1}
		* \left(
			- \suml_{\gamma} \matrix{{cc} 0 & \nabla_K V_{I\beta,K\gamma} \\ \nabla_K V_{K\beta,I\gamma} & 0 }
			* \matrix{{c} C_{Ii,Kk}^{I\gamma} \\ C_{Ii,Kk}^{K\gamma}}
			+ \matrix{{c} \nabla_K (P_{I\beta}|\phi_{Ii}\phi_{Kk}) \\ \nabla_K (P_{K\beta}|\phi_{Ii}\phi_{Kk}) }
		\right)	\, .	\\
\end{array}\end{equation}
Similar to $V$ and $C$, $\nabla V$ and $\nabla C$ also have translational symmetry,
which reduces the computation load and memory consumption.

\subsection{Sparsity exploitation}
\label{sec:sparsity}
As can be seen from Eq.~\eqref{Fexx1}, the calculation of the HFX force involves a summation over numerous items,
each of which is a product of several tensors.
For example,  for a set of atoms $\{A,B,F,G\}$, $\nabla_A C_{AF}^A V_{AB} C_{BG}^B \text{Re}[D_{AB} D_{FG}^*]$
represents an ``item".
When dealing with real systems, the computation times for different items are similar
in magnitude,
assuming that the number of NAOs and ABFs for different elements are comparative.
A key fact is that a large portion of the items have very small values, and
disregarding these items has little effect on the final result, but greatly reduces the computing time. 
In other words, within a generally acceptable error threshold, only a small fraction of the items need to be explicitly calculated.

Therefore, it is highly beneficial to filter out unimportant items in advance
so as to accelerate the calculation.
As alluded to above, each individual item results from a multiplication over 
several two-center tensors:
$C$, $V$, $D$, $\nabla C$, $\nabla V$.
The maximal value of the elements in most of these tensors decays rapidly as the distance between the two atoms increases,
resulting in a large number of items with very small magnitude.

In practice, we employ a two-step procedure to implement the screening idea discussed above. 
In the first step, we filter each individual tensor,
retaining only those with maximal values greater than a pre-specified threshold.
Since each item is the product of several tensors,
if the maximal value of one of the tensors is zero or very tiny,
the resultant item will also be zero or very small, provided there are no extraordinary large numbers from other tensors.
So this item can be safely neglected without incurring significant errors.

In addition to screening the individual tensors, we have also implemented a second screening step to filter out more items, further accelerating the calculation.
This applies to the situation where no individual tensors are small enough to be negligible, yet after being multiplied together, the resultant value of the item is sufficiently small to be discarded. To effectively identify these items, we need to reliably estimate the magnitude of the output
before actually performing the tensor multiplication. Here, we design an algorithm that uses the Cauchy-Schwarz inequality as the estimation criterion.
The Cauchy-Schwarz inequality in matrix form reads
\begin{equation}
	 \left|\mathrm{tr}[AB]\right| \leq \sqrt{\mathrm{tr}[A^+A]} \sqrt{\mathrm{tr}[B^+B]} \, .
\end{equation}
We note that this inequality can be extended to multiple matrices.
For example, the upper limit of the products of three matrices  is
\begin{equation}\begin{array}{ccl}
	|\mathrm{tr}[ABC]|
	&\leq& \sqrt{\mathrm{tr}[A^+A]} \sqrt{\mathrm{tr}[(BC)^+(BC)]}		\\
	&\leq& \sqrt{\mathrm{tr}[A^+A]} \sqrt[4]{\mathrm{tr}[B^+BB^+B]} \sqrt[4]{\mathrm{tr}[C^+CC^+C]}	\, .	\\
\end{array}\end{equation}
The key point of this algorithm is converting a matrix $A$ to a scalar such as $\sqrt{\mathrm{tr}[A^+A]}$.
Before calculating each item by matrix multiplications, it is advantageous to first calculate the scalar products of the individual tensors, from which the upper limit of the item can be quickly estimated.
If the upper limit is smaller than a given threshold,
the corresponding item can be safely discarded.
By applying the two-step screening procedure to the exact-exchange force formula \eqref{Fexx1},
a linear-scaling algorithm is designed, and the force calculation is significantly sped up.
In fact, for different terms in Eq.~\ref{Fexx1}, the best loop structure to implement
the two-step screening algorithm is different.
In Algorithm~\ref{algorithm:loop4}, we illustrate the computation process of the first force 
term of Eq.~\eqref{Fexx1}.
The computation processes for other terms are analogous.

\begin{algorithm}[H]
    \caption{Pseudocode of the filtering process for evaluating the HFX force.}
    \label{algorithm:loop4}
    \begin{algorithmic}[1]
        \ForAll {$A$}
            \ForAll {$B$}
                \If{$\|V_{AB}\| > \theta_V$}
                    \If{$\|D_{AB}\| > \theta_D$}
                        \ForAll {$F$}
                            \If{$\|\nabla_A C_{AF}^A\| > \theta_{\nabla C}$}
                                \ForAll {$G$}
                                    \If{$\|C_{BG}^B\| > \theta_C$}
                                        \If{$\|D_{FG}\| > \theta_D$}
                                            \If{CS$(\nabla_A C_{AF}^A, V_{AB}, C_{BG}^B, D_{AB}, D_{FG}^*) > \theta_{\text{CS}}$}
                                                \State $\bF_A^{\exx1} += \nabla_A C_{AF}^A V_{AB} C_{BG}^B D_{AB} D_{FG}^*$
                                            \EndIf
                                        \EndIf
                                    \EndIf
                                \EndFor
                            \EndIf
                        \EndFor
                    \EndIf
                \EndIf
            \EndFor
        \EndFor
    \end{algorithmic}
\end{algorithm}

\subsection{Parallelization}

In order to use hybrid functionals to relax structures
of large systems containing thousands of atoms in an acceptable amount of time,
it is essential to run these calculations on a supercomputer.
To this end, we have designed a hybrid MPI/OpenMP parallelization scheme in which
the process-level parallelism between nodes is implemented using MPI,
while the thread-level parallelism within node is realized using OpenMP.

As discussed in Sec.~\ref{sec:methods},  
in order to calculate the forces on all atoms in the LRI framework, 
it is necessary to sum over a large number of items, each associated with a set of four atoms (with indices $A,B,F,G$, cf. Eq.~\ref{Fexx1}). This can also be seen from the loop structure given in Algorithm~\ref{algorithm:loop4}, which provides a natural way to assign parallel tasks.
During the initialization stage, the atomic quartets $(A,B,F,G)$ are distributed to different processes,
according to the geometry of the system.
After this, all processes communicate the tensors $C$, $V$, $\nabla C$, $\nabla V$ and $D$ with each other
to retrieve the data required for their assigned computation tasks.
Finally, each process executes an independent loop structure which runs over the allocated atomic quartets on that process.
This is the computationally most intensive step, but
no further communication is needed between different processes.
In this way,  the time waste caused by processes waiting for data input can be effectively avoided.
Additionally, within the loop structure on each process,
items are assigned to individual threads in a dynamic schedule. 
The essential aspects of our parallelization algorithm are twofold.
One is about how to allocate atomic quartets $(A,B,F,G)$ among processes in order to 
achieve load balancing while minimizing memory pressure per process.
The other is on how to communicate large amounts of data between processes, 
while balancing both time and memory consumption.

\begin{figure}[!htbp]
	\includegraphics[width=\textwidth]{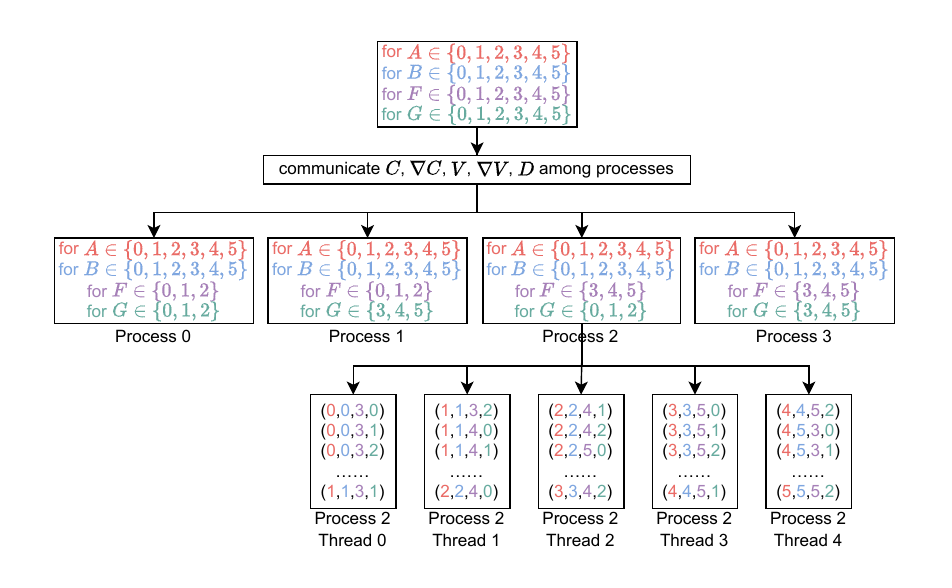}
	\centering
    \caption{Schematic of the MPI/OpenMP parallelization scheme for the HFX force implementation. The test system contains 6 atoms. There are 4 MPI processes and each forks  5 OpenMP threads (for brevity only the threads of Process 2 are shown). Each process is assigned $6^2*3^2$ computation tasks, corresponding to the number of distinct $\langle A, B, F, G \rangle$ (indicated by different colors) atomic quartets,
    with $A,B$ running over all 6 atoms, while $F,G$ running over only three of them. The atomic quartets allocated to each process are further distributed over 5 OpenMP threads. Except for some initial data redistribution, no
    further communication is needed between different MPI processes during the execution through 
    the loop structure. 
	\label{fig:parallel}}
\end{figure}

   The hybrid parallelization scheme described above is schematically illustrated in Fig~\ref{fig:parallel}, using
   a specific example.
    The test system under consideration is very simple, consisting of only six atoms, numbered 0,1,2,3,4,5.
    In our algorithm (cf.~\ref{algorithm:loop4}), each of the atomic indices $A$, $B$, $F$, and $G$ runs through all these six atoms (the filtering can be neglected for such a small system),
    and thus the total number of computation tasks, equivalent to the number of atomic quartets $(A,B,F,G)$, is $6^4$.
    Now assume that there are 20 CPU cores under use when running the program,
    which are divided into 4 processes $\ast$ 5 threads.
    First, the program will distribute the tasks as evenly as possible between the 4 processes.
    In the present case, four processes will be factorized into $4=1*1*2*2$,
     which means that each process will run through all $A$ atoms, all $B$ atoms, half of the $F$ atoms and half of the $G$ atoms,
     as illustrated in Figure~\ref{fig:parallel}.
    Afterward, each process forks five threads.
    These five threads will execute the computation tasks (here the atomic quartets) distributed to that process in order.
    Thus, all CPU cores together compute the entire  list of ``items" forming the HFX force without duplication or omission.

\section{Results}
\label{sec:results}

\subsection{Verification of the correction of the analytical force}
Previously, the functionality for single-point HFX energy calculations had been implemented within the NAO-based ABACUS code \cite{Lin/Ren/He:2020,Lin/Ren/He:2021}. In this work, the implementation is extended to analytical force calculations 
based on the algorithm presented in previous sections. A straightforward way to check the correction
of our analytical force algorithm and implementation is to benchmark against the results obtained from
finite difference (FD) calculations. Such comparison is not complicated
by possible errors due to pseudopotentials, the incompleteness of single-particle and auxiliary 
basis sets, as well as various threshold parameters.

For FD force calculations, the following formula is adopted
\begin{equation}
    F_{\text{FD}}(x)
    = \frac{1}{12\Delta x} [
    -E(x+2\Delta x)+8E(x+\Delta x)-8E(x-\Delta x)+E(x-2\Delta x)]\, ,
\end{equation}
where $x$ denotes the structure at which the force is calculated, and $\Delta x$ is the displacement, here taken to be 0.001~\AA{}. 
This formula is based on the second-order Taylor expansion, which requires four energy points around the structure (here the bond length)
at which the forces are to be determined. Using this formula, the residual error is only $O(\Delta x^4)$. 
To compare with the FD results, we choose two systems: One is an isolated CO molecule and the other is a CO
molecule absorbed on graphene.
In the former case, the forces on the C and O atoms are of opposite sign,
and we only need to look at the force on one of the two atoms (say the C atom).
In the upper part of Table~\ref{tab:CO_force},
we present both the analytical and FD forces on the C
atoms at varying C-O distances. The sign of the force is taken to be positive (negative) at compressed (stretched) bond lengths. 
Table~\ref{tab:CO_force} shows that the differences between the analytical and FD forces are within 0.01 eV/\AA{} for all bond lengths.
The absolute percentage error (APE) is below 1$\%$ when one is away from the equilibrium bond length, and only becomes pronounced
when the equilibrium bond length is approached, because in this regime the magnitude of the force itself gets very small. 
Such a precision is adequate for practical geometry relaxations. We further confirm that the remaining
differences between the analytical and FD forces are due to the numerical integration errors on a uniform real-space grid employed in the ABACUS code. 
The discrepancy can be further reduced by increasing the density of the integration grid.

In the second example, we further examine the forces exerted on the CO molecule upon pressing it towards the graphene surface. 
Specifically, the CO molecule is placed in an upright position with the C atom pointing down to the center of the graphene hexagon. 
When CO is inching towards graphene, we keep the CO bond length fixed at 1.1248 \AA.
In this case, the forces on the C and O atoms are different,
and we look at both of them. In the lower part of Table~\ref{tab:CO_force},
we present the analytical and FD forces on the C and O atoms
as the height of the C atom above the graphene surface is varied. 
At different heights, the magnitude of the forces ranges approximately from 0.3 eV/\AA{} to 5 eV/\AA, 
but the differences between analytical and FD forces are
always at the level of 2 meV/\AA{} or smaller. This example further
confirms the validity of our analytical HFX force evaluation in HDF calculations in more realistic situations.

Similar to the force case,  the correctness of the stress implementation is also checked by comparing the results
obtained from analytical and FD calculations.
In Table~\ref{tab:Si_stress}, we present the analytical and FD stress results for Si FCC crystal structures that
are compressed and stretched from the equilibrium lattice structure.
It can be seen that the differences between the analytical and FD stresses are below 0.1 kbar for all lattice constants,
while the APEs are below 0.1$\%$.
Again, the precision of our analytical stress implementation for hybrid functionals is sufficiently precise for 
cell relaxation in practical calculations.

\begin{table}[!htbp]
\centering
\caption{Comparison of the analytical force $F$ and FD force $F_{\text{FD}}$ for isolated CO molecule (upper part) and CO absorbed on graphene (lower part). 
The first column (Dist) contains the distances between the C and O atoms for the isolated CO molecule and the height of C atom above
the graphene for the absorbed CO case. The APD is given by $|\Delta F/F_{\text{FD}}|$.}
\label{tab:CO_force}
\begin{tabular}{cccccc}\hline
    Dist(\AA) & atom & $F$ (eV/\AA) & $F_{\text{FD}}$ (eV/\AA) & $\Delta F$ (eV/\AA) & APD (\%)     \\ 
    \hline  
    \multicolumn{6}{c}{Isolated CO} \\ [1.0 ex] 
    0.9000	& C & 66.3323 	& 66.3372 	& -0.0048 	& 0.01\%		\\
	1.0000	& C & 25.1073 	& 25.1098 	& -0.0025 	& 0.01\%		\\
	1.1000	& C & 3.4735 	& 3.4801 	& -0.0066 	& 0.19\%		\\
	1.1248	& C & 0.0225 	& 0.0271 	& -0.0046 	& 17.08\%		\\
	1.1500	& C & -2.9161 	& -2.9143 	& -0.0018 	& 0.06\%		\\
	1.2000	& C & -7.3498 	& -7.3427 	& -0.0071 	& 0.10\%		\\
	1.3000	& C & -12.2985 	& -12.2901 	& -0.0084 	& 0.07\%		\\ [1ex] \hline
    \multicolumn{6}{c}{CO@graphene}\\ [1.0 ex] 
   1.50   &C&	1.4903 	&	1.4902 	&	0.0001 	&	0.01\%	\\
        &O&    1.4155 	&	1.4132 	&	0.0023 	&	0.16\% \\
	1.75   &C&	5.6559 	&	5.6536 	&	0.0024 	&	0.04\%	\\
        &O&    1.1214 	&	1.1238 	&	-0.0024 	&	0.22\% \\
	2.00   &C&	4.9648 	&	4.9645 	&	0.0004 	&	0.01\%	\\
        &O&    0.8607 	&	0.8589 	&	0.0018 	&	0.21\% \\
	2.25   &C&	2.9055 	&	2.9047 	&	0.0007 	&	0.03\%	\\
        &O&    0.5271 	&	0.5284 	&	-0.0013 	&	0.25\% \\
	2.50   &C&	1.4837 	&	1.4815 	&	0.0021 	&	0.14\%	\\
        &O&    0.2878 	&	0.2899 	&	-0.0021 	&	0.73\% \\\hline
\end{tabular}
\end{table}

\begin{table}[!htbp]
\centering
\caption{Comparison of the analytical stress $\sigma$ and the FD stress $\sigma_{\text{FD}}$ for compressed
and stretched Si FCC crystal structures.
The APD is given by $|\Delta \sigma/\sigma_{\text{FD}}|$. }
\label{tab:Si_stress}
\begin{tabular}{ccccc}\hline
    lattice constant (\AA) & $\sigma$ (kbar) & $\sigma_{\text{FD}}$ (kbar) & $\Delta \sigma$ (kbar) & APD (\%)     \\\hline
5.0 	& 1299.8096 	& 1299.8077 	& 0.0019 	& 0.0001\%		\\
5.1 	& 915.5895 	& 915.6040 	& -0.0144 	& 0.0016\%		\\
5.2 	& 596.1901 	& 596.1853 	& 0.0047 	& 0.0008\%		\\
5.3 	& 346.3322 	& 346.3597 	& -0.0275 	& 0.0079\%		\\
5.4 	& 121.5473 	& 121.5920 	& -0.0447 	& 0.0368\%		\\
5.5 	& -70.9570 	& -70.9547 	& -0.0024 	& 0.0033\%		\\
5.6 	& -194.2220 	& -194.2309 	& 0.0089 	& 0.0046\%		\\
5.7 	& -280.7348 	& -280.7123 	& -0.0225 	& 0.0080\%		\\
5.8 	& -355.8014 	& -355.8134 	& 0.0120 	& 0.0034\%		\\
5.9 	& -406.2139 	& -406.2553 	& 0.0415 	& 0.0102\%		\\
6.0 	& -447.3744 	& -447.4254 	& 0.0510 	& 0.0114\%		\\
\hline
\end{tabular}
\end{table}

\subsection{Effect of filtering individual tensors}

The key aspect of our algorithm is that we can efficiently filter out small components of the
tensors that are contracted to yield the analytical HFX force, enabling us to relax structures of large systems using HDFs. 
In Sec.~\ref{sec:methods}, we have described in detail how to exploit the sparsity of these tensors to
achieve linear scaling in HFX force calculations. In this work, we will check the actual performance of our algorithm.
Obviously, if more elements of the tensors are filtered out,
the computational efficiency will be improved, but the truncation error might also increase. 
Therefore, there is a trade-off between computational speed and numerical accuracy.

In our implementation, the screening is carried out in terms of small dense blocks of the tensor, each associated with an atomic pair.
When the maximum matrix element of such a block is below than a given threshold value,
the entire block will be discarded.
By tuning the screening threshold, 
the influence of these tensors on the accuracy and speed of the calculation can be quantitatively measured.

Here, to be specific, we choose the Si crystal as a prototypical test sample,
with each unit cell containing only two atoms
that can be displaced from their equilibrium positions. 
Here we use $8\times 8 \times 8$ \bfk~grid for the BZ sampling, corresponding to a $8\times 8 \times 8$ BvK supercell in real space. 
In this system setting, on the one hand,
there are enough tensor elements and the ``items", with values spreading over a wide range, 
to test our screening algorithm.
On the other hand, each unit cell contains only two atoms, when displaced,
resulting in only one pair of forces with the same magnitude and opposite directions.
This type of system is therefore suitable for testing the screening effect.

\begin{figure}[!htbp]
    \includegraphics[width=0.3\textwidth]{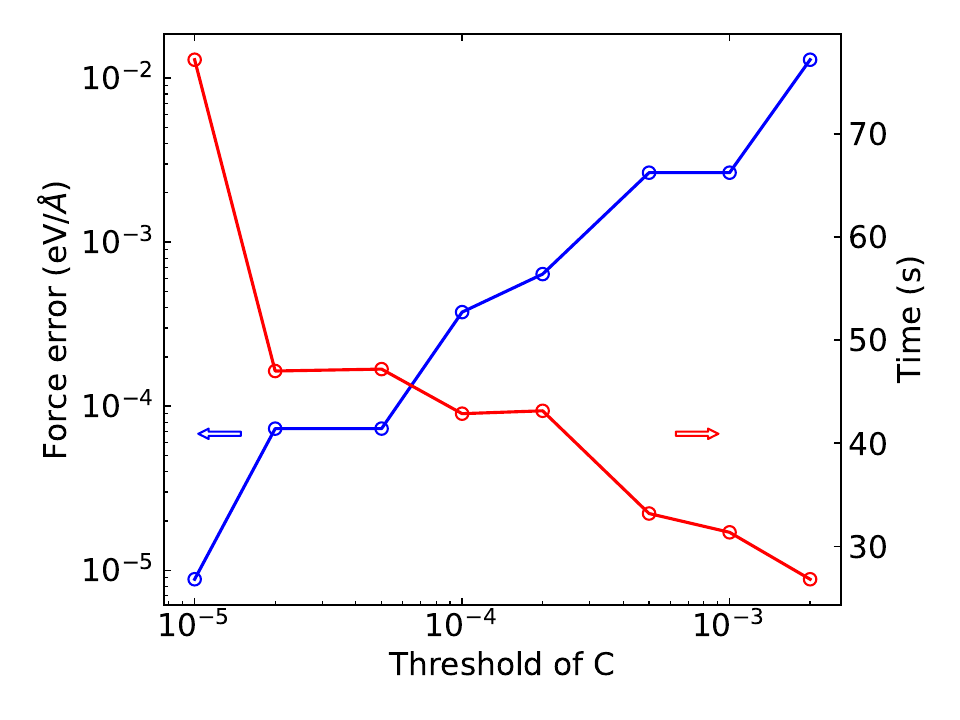}
    \includegraphics[width=0.3\textwidth]{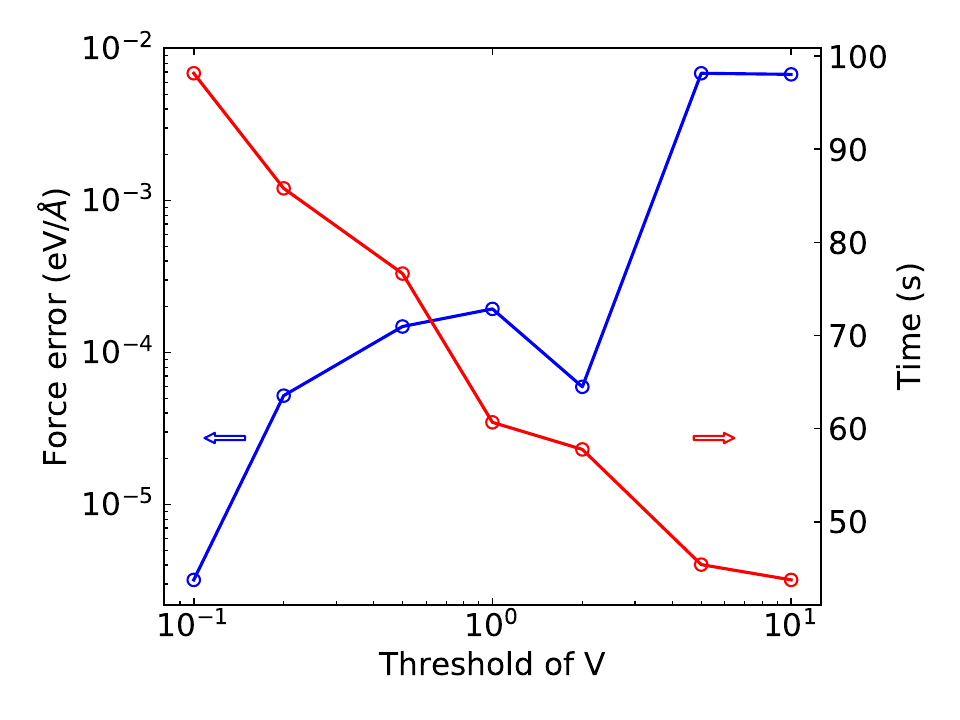}
    \includegraphics[width=0.3\textwidth]{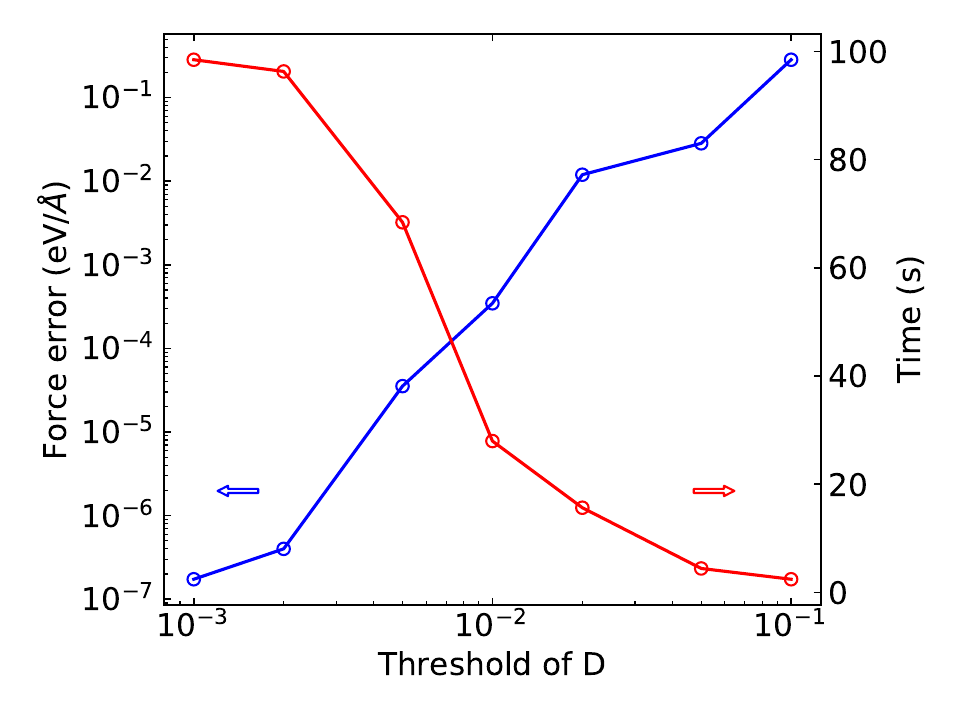}
    \includegraphics[width=0.3\textwidth]{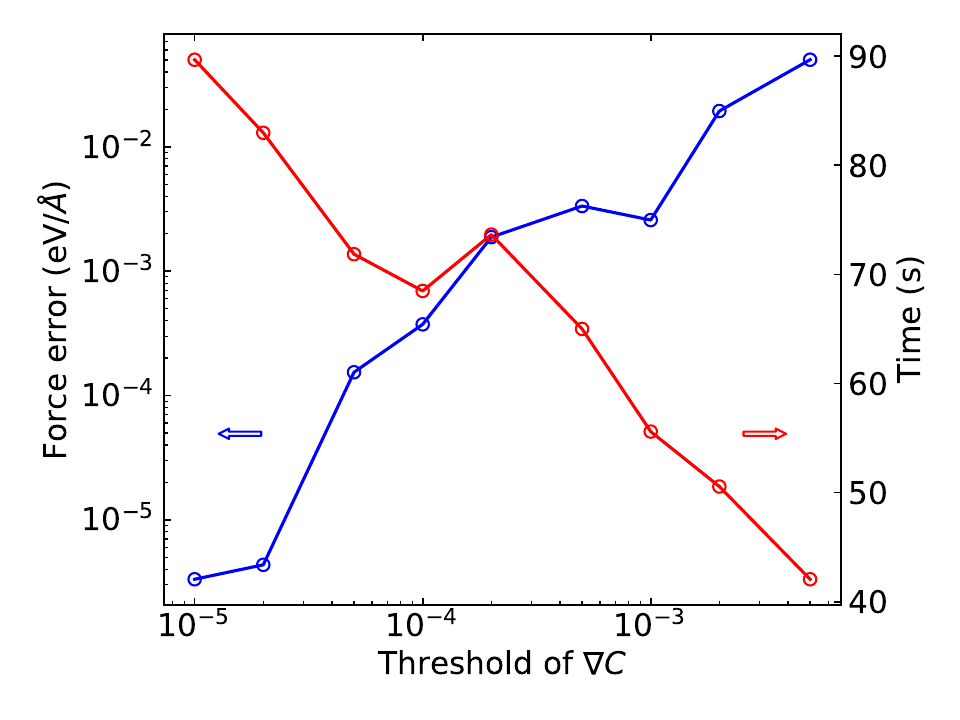}
    \includegraphics[width=0.3\textwidth]{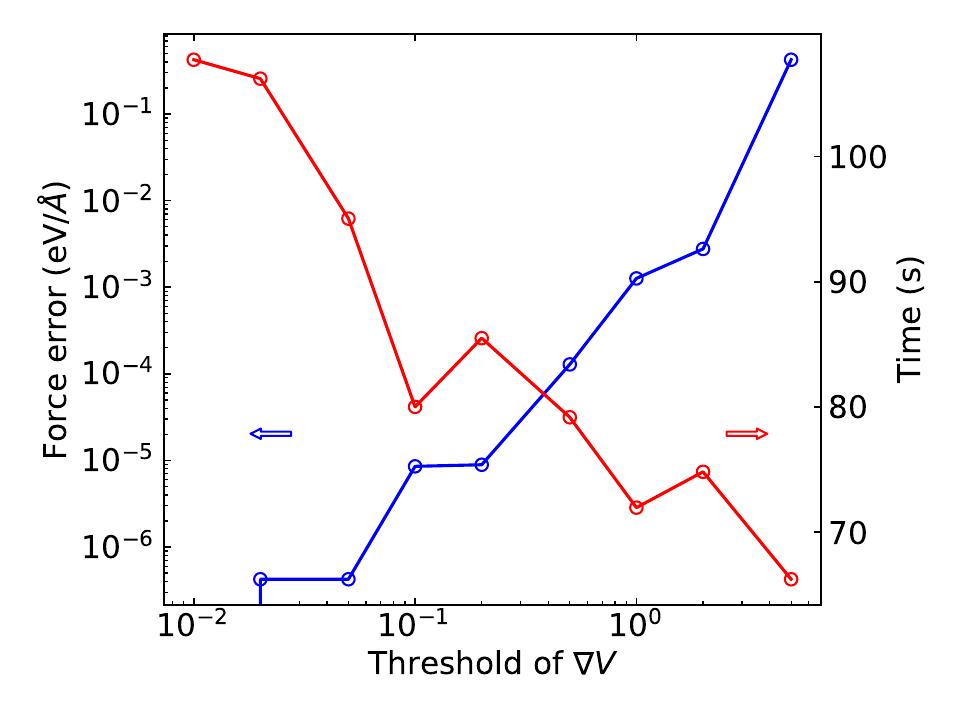}
	\centering
    \caption{Computation timings of the HFX force calculations as a function of the threshold parameters in prescreening the individual $C$, $V$, $D$, $\nabla C$ and $\nabla V$ tensors.
    The Si crystal with  $8\times 8 \times 8$ \bfk~grid is used as the test system.
    In all panels, the accompanying force errors are plotted as an indicator of the numerical accuracy of the calculations.}
	\label{fig:thr_tensor}
\end{figure}

As can be seen from Eq.~\ref{Fexx1}, the HFX force depends on five tensors: 
$C$, $V$, $D$, $\nabla C$ and $\nabla V$. In Figure~\ref{fig:thr_tensor}, 
we check separately the influence of the screening process on each individual tensor.
Here, the calculation errors presented are defined as the difference
between the forces obtained with the screening and those without invoking any screening,
while the computation time is the actual wall-clock time for evaluating the HFX force.
We note that, in each panel of Figure~\ref{fig:thr_tensor}, 
when testing the screening effect of one tensor, the screening thresholds of
all other tensors are set to zero.

As is expected, the incurred errors and the time reductions resulting from screening all five tensors follow similar patterns: 
As the screening threshold raises, the calculation error increases, while the corresponding calculation time decreases. One may note that variation of the error and the computation time as
a function of the screening threshold is not completely monotonic, and this is due to the
delicate error compensation and the complicated runtime environment when performing the calculations.

The key question here is to choose an appropriate set of threshold parameters that can speed up the calculation while still maintaining adequate numerical accuracy.
The screening threshold can be chosen following the following recipe:
According to the maximum force error that is allowed in actual calculations,
the highest threshold value that satisfies the accuracy requirement is selected to achieve the fastest calculation speed. We do this not for a single material, but for a set of 21 insulators and semiconductors
as listed in Table~\ref{tab:thr_crystal}.
The detailed test results of these 21 crystals are presented in Fig.~S1 of the Supporting Information (SI).
Specifically, if the tolerance for the force error is set to be $10^{-3}$ eV/\AA, 
the largest acceptable threshold values for each of the five tensors are given in Table~\ref{tab:thr_default}. The actual procedure to determine the default set of
parameters is illustrated in Fig.~S1 of the SI.
With such threshold settings, force errors are below
$10^{-3}$ eV/\AA{} for all materials, but the average error across the entire set of materials
is even smaller, as also shown in Table~\ref{tab:thr_default}. Finally, when all five tensors are screened simultaneously with the aforementioned thresholds, the errors in the HFX force calculations for individual materials are presented in Table~\ref{tab:thr_crystal}.
The table also shows the time reduction in percentage averaged over all materials when each tensor is individually screened. It should be noted that the actual time savings depend on the system size. For the tests in this section, an $8\times 8 \times 8$ BvK supercell is considered. 
The time reduction will be more drastic as the system size grows. 
\begin{table}[!htbp]
	\centering
        \caption{The largest allowable threshold values for screening individual tensors. Under such choices,
        the incurred force errors are all below $10^{-3}$ eV/\AA {}  for a set of 21 crystals. The average errors in force and the time reductions resulting from screening individual tensors are presented. 
        }
	\begin{tabular}{cccc}\hline
		tensor	& threshold	& average error of force (eV/\AA)	& average relative time	\\\hline
		$C$			& $10^{-4}$	& $7.53*10^{-5}$	& 58.64\%	\\
		$V$			& $10^{+0}$	& $1.02*10^{-4}$	& 59.15\%	\\
		$D$			& $10^{-3}$	& $3.26*10^{-5}$	& 60.18\%	\\
		$\nabla C$	& $10^{-4}$	& $1.31*10^{-4}$	& 76.77\%	\\
		$\nabla V$	& $10^{-1}$	& $1.05*10^{-5}$	& 83.36\%	\\\hline
	\end{tabular}
	\label{tab:thr_default}
\end{table}

\begin{table}[!htbp]
	\centering
   \caption{The errors in the HFX force and the time savings for 21 crystalline materials 
   when screening simultaneously all five tensors with the choices of threshold parameters listed in Table~\ref{tab:thr_default}.
        }
	\begin{tabular}{cccc}\hline
		crystal	& error of force(eV/\AA)	& relative time	\\\hline
		AlAs	& $7.23*10^{-4}$	& 9.41\%	\\
		AlP 	& $2.12*10^{-4}$	& 8.90\%	\\
		AlSb	& $1.32*10^{-4}$	& 30.59\%	\\
		BN  	& $1.63*10^{-4}$	& 7.60\%	\\
		BP  	& $4.69*10^{-4}$	& 18.35\%	\\
		C   	& $4.96*10^{-4}$	& 7.89\%	\\
		CdTe	& $1.10*10^{-4}$	& 15.05\%	\\
		GaAs	& $2.00*10^{-5}$	& 27.70\%	\\
		GaN 	& $4.30*10^{-5}$	& 15.20\%	\\
		GaP 	& $1.67*10^{-5}$	& 29.51\%	\\
		GaSb	& $2.24*10^{-4}$	& 39.63\%	\\
		InP 	& $3.03*10^{-4}$	& 10.00\%	\\
		LiF 	& $3.63*10^{-4}$	& 4.32\%	\\
		MgS 	& $1.83*10^{-4}$	& 9.90\%	\\
		NaCl	& $3.27*10^{-6}$	& 13.73\%	\\
		Si  	& $6.18*10^{-4}$	& 12.17\%	\\
		SiC 	& $9.75*10^{-5}$	& 10.60\%	\\
		ZnO 	& $2.69*10^{-5}$	& 11.04\%	\\
		ZnS 	& $1.30*10^{-4}$	& 10.17\%	\\
		ZnSe	& $8.07*10^{-5}$	& 9.09\%	\\
		ZnTe	& $3.90*10^{-4}$	& 12.16\%	\\\hline
	\end{tabular}
	\label{tab:thr_crystal}
\end{table}

\subsection{Effect of the screening based on the Cauchy-Schwarz inequality}

As discussed in Sec.~\ref{sec:sparsity}, after filtering out small elements of individual 
tensors, we introduce a second-step screening based on the Cauchy-Schwarz inequality. 
This can further filter out small ``items", for which the magnitude of the individual tensors
is still larger than the pre-given thresholds and thus escaped the first-step screening.
Similar to the discussion in the previous subsection, the error incurred and the calculation
times can be measured by varying the threshold values.

\begin{figure}[!htbp]
    \includegraphics[width=0.48\textwidth]{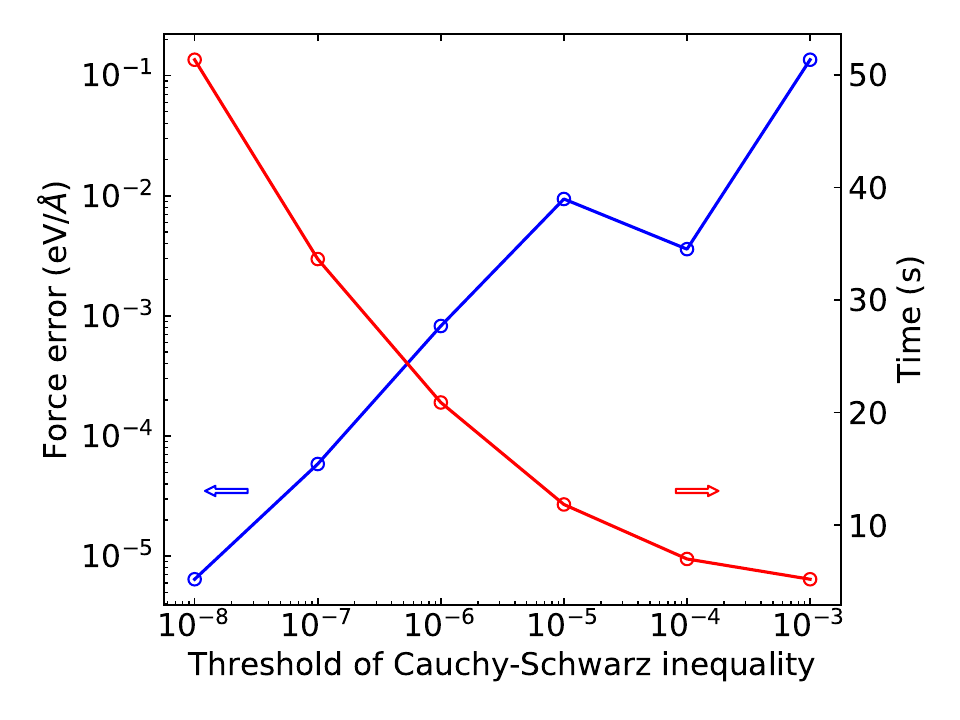}
    \includegraphics[width=0.48\textwidth]{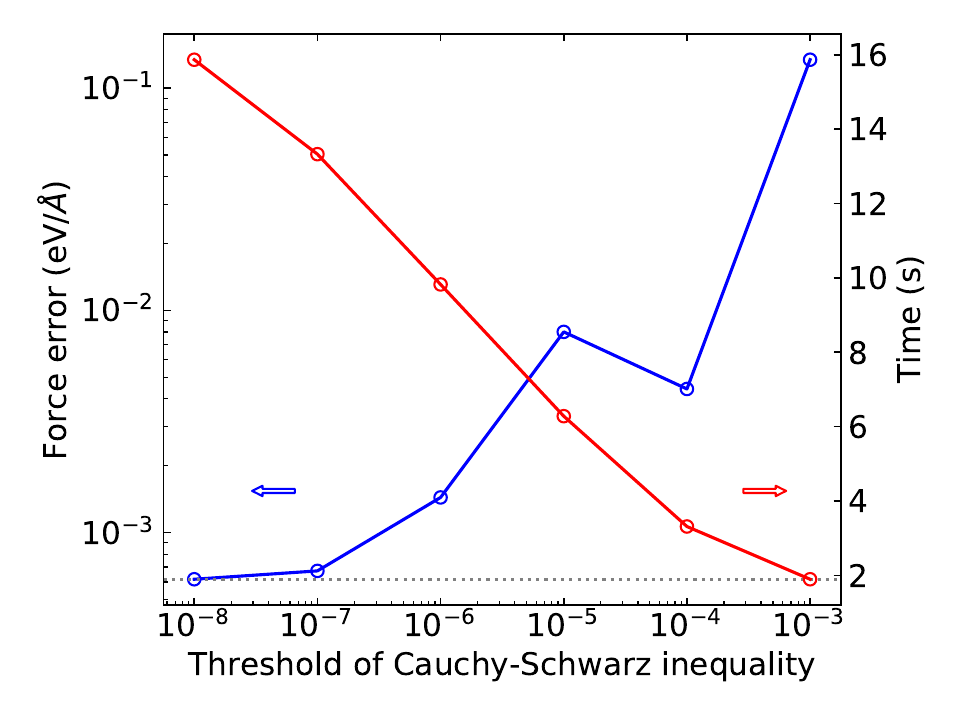}
	\centering
    \caption{Computation times as a function of the threshold parameter of the Cauchy-Schwarz screening.
    The test system is also the Si crystal with $8\times 8 \times 8$ \bfk~grid.
    The left panel presents the results which are purely based on the Cauchy-Schwarz inequality 
    while the right panel starts with tensor-based pre-screening
    and proceeds with a further screening based on Cauchy-Schwarz inequality.
    With the decrease of threshold values, the force error of in the left panel approaches 0, 
    while in the right panel, the force error approaches $6.18*10^{-4}$ eV/\AA, due to the
    prescreening of individual tensors (cf. Table~\ref{tab:thr_crystal}).
    }
	\label{fig:thr_cs}
\end{figure}

Figure~\ref{fig:thr_cs} presents two sets of test results. 
The first (left panel) invokes only screening based on the Cauchy-Schwarz inequality
without any tensor-based screening, 
while the second (right panel) performs individual tensor-based screenings beforehand
according to the threshold values listed in Table~\ref{tab:thr_default}. 
From Figure~\ref{fig:thr_cs}, 
it can be observed that the Cauchy-Schwarz inequality can effectively further reduce the computational cost by filtering out small ``items" beyond the tensor-based screening.
Additionally, comparing the two panels in Figure~\ref{fig:thr_cs} reveals that 
employing the Cauchy-Schwarz inequality alone without pre-screening the individual tensors (left panel)
is not as effective as the combined two-step screening approach (right panel), 
suggesting that it cannot completely replace the individual tensor-based screening.
As presented in Fig.~2 of the SI,
the test results on 21 crystals show that for a given tolerance $10^{-3}$ eV/\AA,
it's appropriate to set the threshold of Cauchy-Schwarz inequality $\theta_{\text{CS}}$ as $10^{-7}$.

\subsection{Scaling behavior with respect to system size}

In previous sections, we examined the efficacy of our
screening algorithm in simple systems, with a small unit cell and a fixed number of $\bfk$ points. 
Here we further check the scaling behavior of the computational time with respect to system size. 

Again, utilizing the Si crystal as the test system, we check how
the computational time for HDF (here the HSE functional) force calculations 
grows with respect to the supercell size. 
In these calculations, a $1\times 1 \times 1$ $\Gamma$-only $\bfk$ grid is used.
For the one-electron basis set, the NAO-DZP ($2s2p1d$) generated by the DPSI method \cite{PhysRevB.103.235131}
is employed, while the corresponding ABF basis is generated by the principal component analysis (PCA) method \cite{Lin/Ren/He:2020} with threshold $\theta_{\text{PCA}} = 10^{-4}$, ending up with an ABF set of $5s5p5d2f1g$.
The cutoff radius of all the basis functions is 7 Bohr. Furthermore, the screening threshold values for individual tensors are specified in Table~\ref{tab:thr_default}, and a value of $\theta_{\text{CS}}=10^{-7}$ is used for the Cauchy-Schwarz screening. With such
threshold settings, the error of force is $6.73*10^{-4}$ eV/\AA, which satisfies the
accuracy requirement for practical calculations. 

\begin{figure}[!htbp]
	\includegraphics[width=0.7\textwidth]{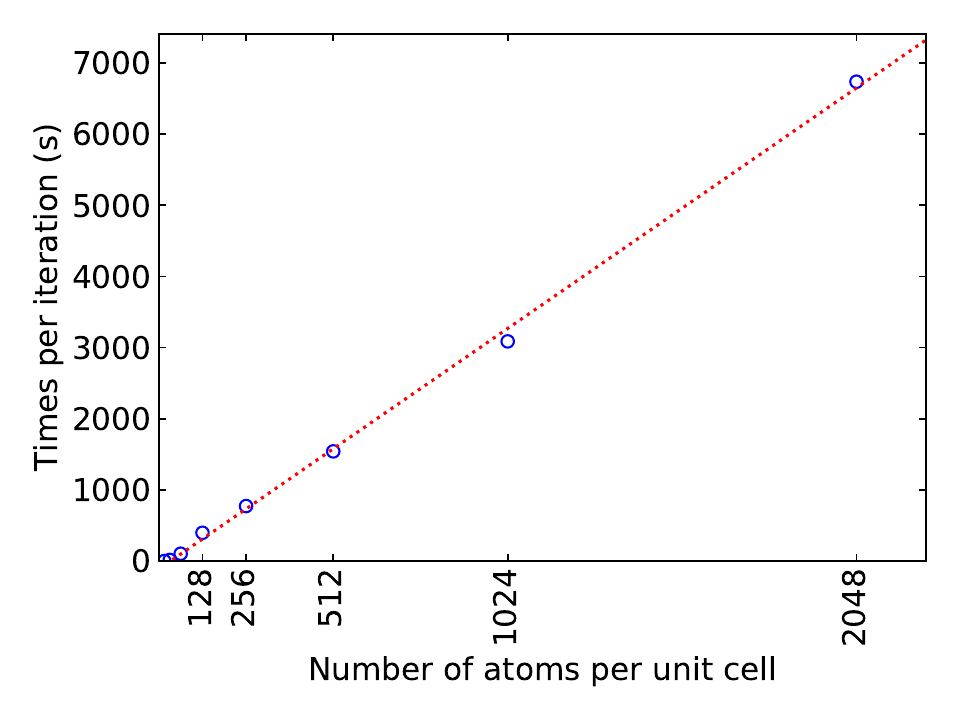}
	\centering	\caption{Computational time versus the supercell size for the Si crystal.
   The calculations were performed on a single node equipped with two Intel Gold 6330 CPUs, each with 28 cores. }
 \label{fig:scaling_atoms}
\end{figure}

As shown in Figure~\ref{fig:scaling_atoms}, the computational time for calculating the HDF forces 
increases linearly with the number of atoms contained in the supercell. 
The dashed line in the figure represents a linear fit of the data beyond 128 atoms. 
This fitting exhibits a linear relationship between the computational time $t$ and
the number of atoms $N$ in the system, 
\begin{equation}
  t=3.30 N - 111.64
\end{equation}
in unit of seconds, with a confidence level  $R^2=0.9981$.
When HDFs are used for geometry relaxations and molecular dynamics, 
The mild linear-scaling increase in the computational cost is the key for such
calculations to be feasible for large systems.

As discussed previously, our HDF implementation achieves linear scaling by sufficiently exploiting the sparsity of the RI expansion coefficients and the short-range nature of 
the density matrix. 
Obviously, the sparsity only becomes significant when the system size exceeds a certain range. 
From our tests, it is observed that the linear scaling already sets in 
when the supercell size exceeds 128 atoms, 
corresponding to 4 times of the primitive cells of Si along each dimension. This demonstrates the
high efficacy of our linear-scaling algorithm to deal with large systems.

\subsection{Parallel efficiency}

As mentioned earlier, our HFX force implementation is massively parallelized
based on a hybrid MPI/OpenMP scheme.
As one of the most critical metrics for measuring parallel efficiency,
here we present the test results of strong scalability of our implementation.
That is, the computing time is recorded by increasing the number of CPU cores utilized while keeping the system size fixed. 

\begin{figure}[!htbp]
    \includegraphics[width=0.48\textwidth]{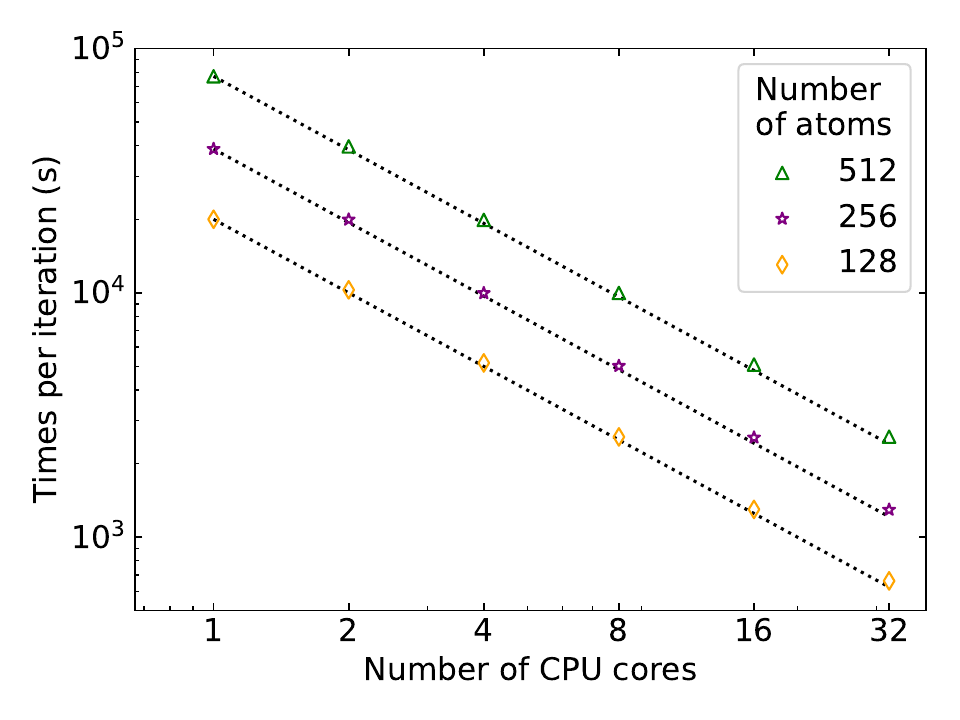}
    \centering
    \caption{Strong scalability of the computational time of one iteration of the HSE
 force calculations with respect to the number of OpenMP threads (equal to the number of CPU cores). 
 Different lines correspond to different supercell size of the Si crystal structure.}
    \label{fig:scaling_openmp}
\end{figure}

We first test the parallel efficiency with respect to the OpenMP threads.
In this case, the calculations were done by employing only one MPI process, but
different numbers of OpenMP threads. 
Figure~\ref{fig:scaling_openmp} shows that the thread scalability of our implementation is remarkably good, 
with nearly linear speedup across different system sizes.
Upon each doubling of the thread count, the computational time is nearly halved.
For example, for the 512-atom system, when going from a single thread to 32,
the calculation was sped up 29.87 times, achieving a parallel efficiency of 93.35\%.

\begin{figure}[!htbp]
    \includegraphics[width=0.6\textwidth]{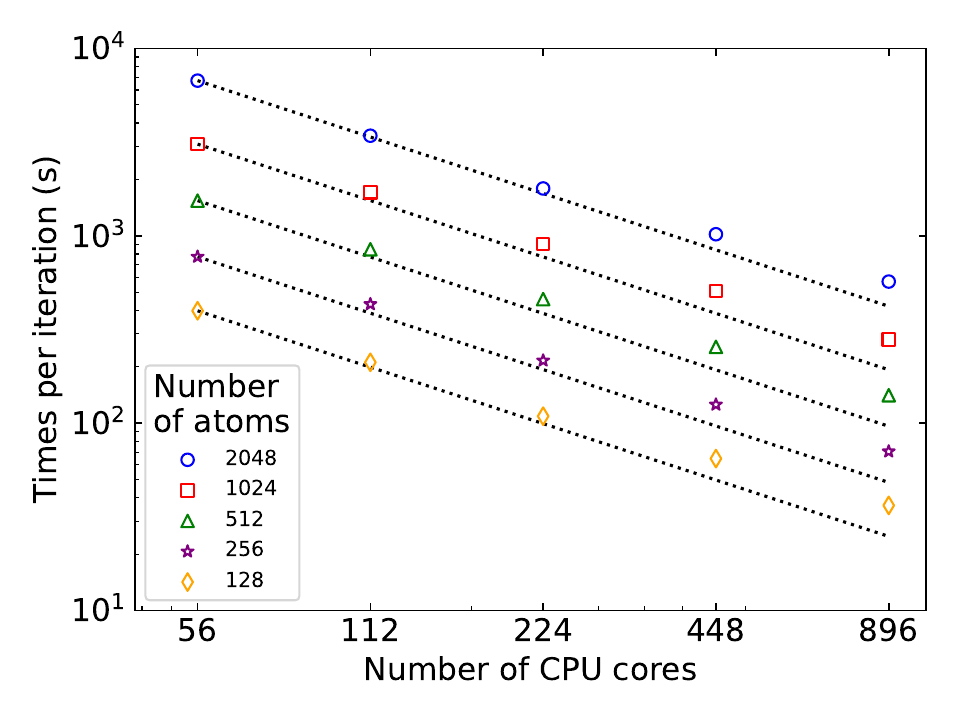}
    \centering
    \caption{Strong scaling of the computational time with respect to the number of
 CPU cores (the MPI processes) in the hybrid MPI/OpenMP parallelization scheme. In these calculations, 
 each MPI process forks 56 OpenMP threads. Different lines correspond to different supercell sizes of
 the Si crystal system.}
    \label{fig:scaling_mpi}
\end{figure}

Next, we test the parallel efficiency with respect to the MPI processes. 
In our algorithm design, MPI is primarily used for communications between nodes, 
and thus the most efficient approach is to initiate one process per node. 
On the supercomputer platform used in this study, 
a compute node is equipped with two Intel Gold 6330 CPUs, each with 28 cores, and thus each process can 
fork 56 threads.
As shown in Figure~\ref{fig:scaling_mpi}, the computational time gradually decreases with increasing 
number of processes.
However, the parallelization with respect to MPI processes has not reached the ideal linear speedup. 
For example, by fitting the data of the 2048-atom system,
the computing time scales as $\left(1/N_{proc}\right)^{0.868}$ with $N_{proc}$ being the number of MPI processes. 
When going from one process (56 cores) to 16 processes (896 cores), one achieves a speedup of 11.81, with a parallel efficiency of 73.83.
The deviation from a linear speedup is mainly because distribution of computational tasks among processes is 
not yet fully balanced.
In the subsequent work, further optimization of the algorithm and code
will be carried out with the objective of approaching linear speedup. 

\subsection{Application to the structure relaxation of halide perovskites}
Halide perovskites (HPs), recognized for their low cost\cite{Song/etal:2017}, excellent carrier transport characteristics\cite{Oga/etal:2014,Samuel/etal:2013,Guichuan/etal:2013,dong/etal:2015,shi/etal:2015}, and outstanding optoelectronic performance\cite{Wolf/etal:2014}, have been widely applied in solar cells\cite{Liu/etal:2024,Li/etal:2024,Chen/etal:2024}, light-emitting diodes\cite{Fakharuddin/etal:2022,Veldhuis/etal:2016,Lin/etal:2018}, detectors\cite{Li/etal:2024,He/etal:2021}, spintronic devices\cite{Kepenekian/etal:2017,Long/etal:2020}, and other technologies\cite{Hwang/etal:2017,Wu/etal:2019}. The exceptional semiconductor behavior of this type of materials is closely related to their soft lattices, with factors such as symmetry, octahedral tilting, and distortions playing pivotal roles in shaping their electronic structure and other properties\cite{Huang/etal:2022, Swift/Lyons:2023,Wiktor/Fransson/Dominik:2023, Balvanz/etal:2024, Fabini/etal:2024, Hylton/Colin/Remsing:2024}.

Emphanisis, a phenomenon of local symmetry breaking driven by lone-pair electrons upon heating, has been observed in  HPs with a formula of $ABX_3$\cite{Fabini/etal:2016,Huang/etal:2022,Maria/etal/2024}. Specifically, the $s^2$ lone-pair stereochemistry of the $B$-site metal cation induces dynamic off-center displacements of the $B$ ions. DFT has been employed to investigate the relationship between lattice distortion and off-centering\cite{Fabini/etal:2016,Huang/etal:2022,Gu/etal/2023,Fabini/etal:2024}. It was found in Ref.~\citenum{Swift/Lyons:2023} that semi-local functionals tend to underestimate the stereochemical effects of $B$-site $s^2$ lone pairs, while hybrid functionals (e.g. HSE) that incorporate spin–orbit coupling (SOC) are necessary to accurately capture the local asymmetry and its impact on the electronic structure of HPs. 

\begin{figure}[!htbp]
	\includegraphics[width=\textwidth]{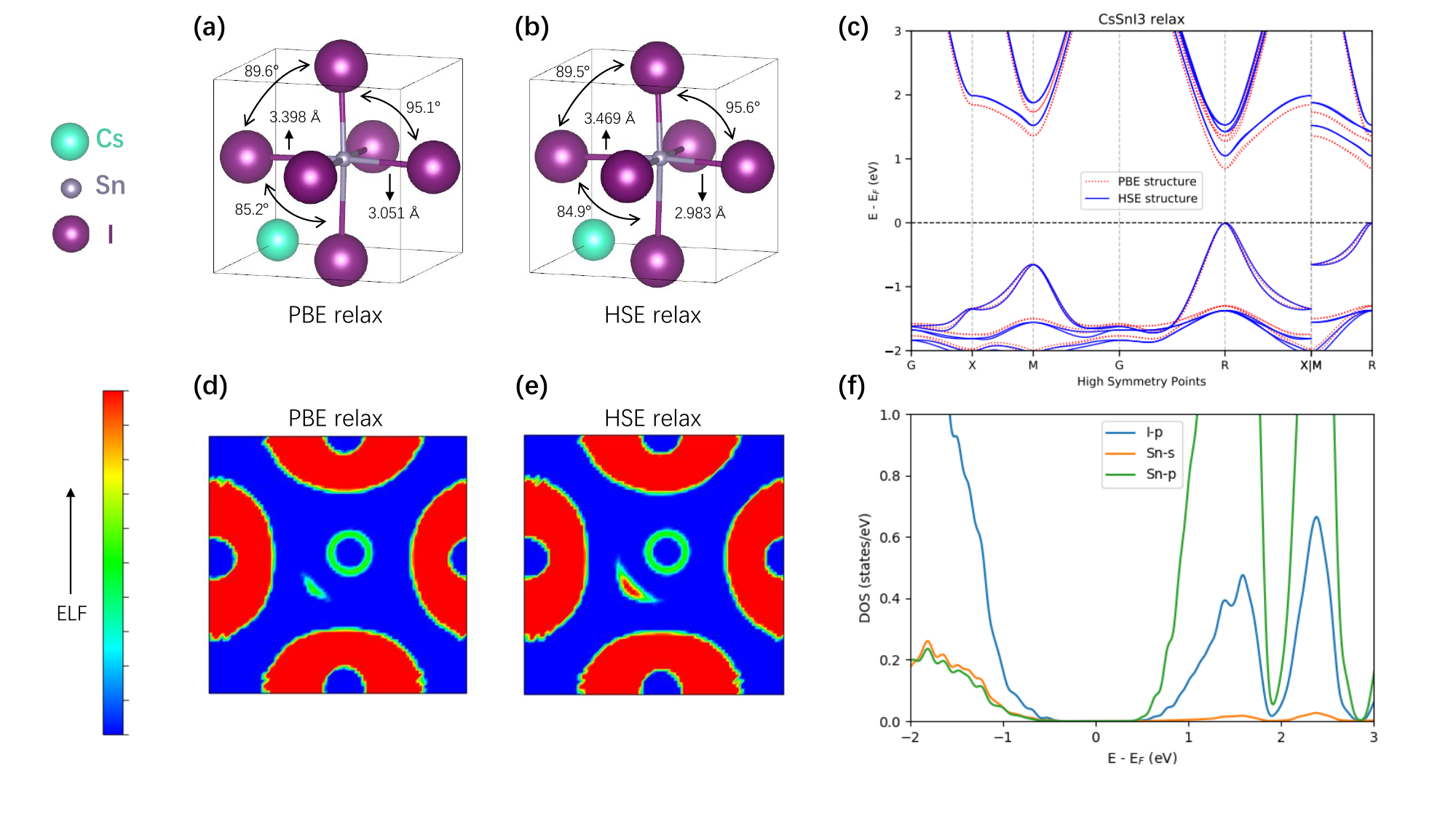}
	\centering
    \caption{(a)-(b) Cubic CsSnI$_3$ relaxed with PBE and HSE under fixed lattice parameters, with bond lengths and bond angles marked. The corresponding HSE+SOC band structure is depicted in Figure (c). (d)-(e) ELFs on the (001) plane for the PBE and HSE-optimized structures, respectively. (f) HSE+SOC PDOS based on the HSE structure.
 	\label{fig:HDP_relax}}
\end{figure}

In this study, we investigate the $s^2$ lone-pair activity in the all-inorganic perovskite CsSnI$_3$ and its Ge-doped counterpart, employing both the PBE and HSE functionals. For CsSnI$_3$, we adopt a cubic unit cell, first with fixed lattice parameters (Table~\ref{tab:HDP_111}) through the optimization process, allowing only the atomic positions to relax. The optimized bond lengths and bond angles for both functionals are presented in Figure~\ref{fig:HDP_relax} (a) and (b). The differences in bond lengths are on the order of $10^{-2}$ \AA, while the differences in bond angles are on the order of $10^{-1}$ degrees.

\begin{table}[!htbp]
    \centering
    \caption{Comparison of octahedral distortion parameters for the atomic position-relaxed and fully relaxed CsSnI$_3$ unit cell, optimized using PBE and HSE. Here, the lattice parameters are $a=b=c$ and $\alpha=\beta=\gamma$.}
    \label{tab:HDP_111}
    \begin{tabular}{c | c | c | c | c | c | c }
        \toprule 
            Method & $a$ & $\alpha$ & $d^0_\text{Sn-I}$(\AA) & $\sigma_1^2$ (deg$^2$) & $\sigma_2^2$ (deg$^2$) & $\lambda$ \\
        \midrule 
            PBE relax & 6.437 & 90 & 3.224 & 17.6 & 72.8 & 1.003\\ 
            HSE relax & 6.437 & 90 & 3.226 & 21.1 & 87.2 & 1.006 \\
            PBE cell-relax & 6.304 & 89.6 & 3.156 & 12.0 & 48.9 & 1.001 \\
            HSE cell-relax & 6.280 & 89.3 & 3.147 & 20.4 & 82.7 & 0.997\\
        \bottomrule 
    \end{tabular}
\end{table}

Moreover, the octahedral distortion parameters for the relaxed structures are summarized in Table~\ref{tab:HDP_111}. These parameters are defined following the methodology described in Ref.~\citenum{Huang/etal:2022}, including the bond angle variance ($\sigma_1^2=\frac{1}{11}\sum_{i=1}^{12}(\theta_i-90)^2$ and $\sigma_2^2=\frac{1}{2}\sum_{i=1}^{3}(\eta_i-180)^2$), and bond length quadratic elongation ($\braket{\lambda}=\frac{1}{6}\sum_{i=1}^6(\frac{d_i}{d_0})^2$). Here, $\theta_i$ represents the I-Sn-I bond angle between neighboring Sn-I bonds, $\eta_i$ denotes I-Sn-I bond angle between non-neighboring Sn-I bonds, $d_i$ is the Sn-I bond length, and $d_0$ is the mean Sn-I bond distance.

We further conduct electron localization function (ELF) calculations based on the PBE and HSE-optimized structures. Crescent-shaped ELFs are both observed around the Sn ions, indicating octahedral distortion. HSE predicts higher crescent-shaped ELF values (in red), suggesting that the Sn 5$s^2$ electronic states are more localized in the HSE case, which may be associated with the mitigated delocalization error. 

Figure~\ref{fig:HDP_relax}(c) shows the HSE+SOC band structures calculated based on PBE and HSE relaxed structures. It can be observed that the difference in the Sn off-centering between the two structures leads to shifts in the band edges. This is attributed to changes in the hybridization of the Sn s and p orbitals with the I p orbital, caused by the elongation (or shortening) of the bond lengths, as revealed by the partial density of states (PDOS) analysis (Figure~\ref{fig:HDP_relax} (f)). PBE underestimates Sn $s^2$ lone-pair activity, resulting in a smaller HSE+SOC band gap (0.85 eV) compared to the one (1.04 eV) obtained from the HSE structure.

\begin{figure}[!htbp]
	\includegraphics[width=\textwidth]{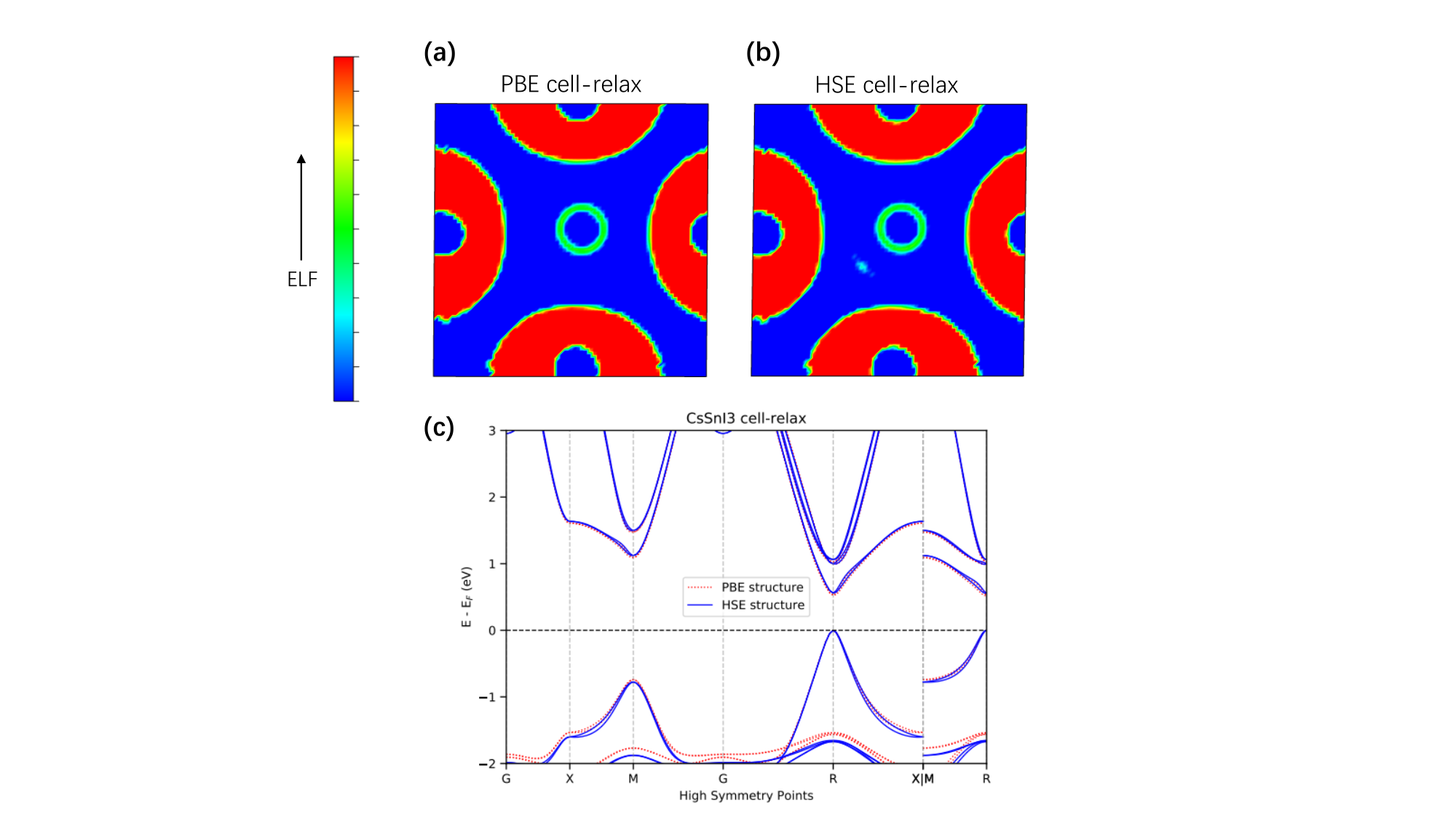}
	\centering
    \caption{(a)-(b) ELFs on the (001) plane for the fully relaxed PBE and HSE CsSnI$_3$ unit cells, with both atomic positions and lattice parameters optimized, respectively. (c) HSE+SOC band structures calculated with PBE and
    HSE structures. 
	\label{fig:HDP_cell-relax}}
\end{figure}

We further perform full relaxation of the unit cell using PBE and HSE, allowing for variations in both atomic positions and lattice parameters. According to Table~\ref{tab:HDP_111}, compared to the structures optimized with fixed lattice constants, the fully optimized structures using PBE and HSE exhibit lattice compression, resulting in a quasi-cubic structure with reduced local distortion. A further comparison of the ELF analyses in Figure~\ref{fig:HDP_cell-relax} (a) and (b) show that, after full optimization with PBE and HSE, the stereochemical effect of the Sn $s^2$ lone pairs is weakened, as no significantly high ELF values are observed near Sn ion. Therefore, lattice strain and lone pair expression induce counteractive effects, as distortion requires sufficient space to exist\cite{Swift/Lyons:2023}. The small differences in lattice parameters and local structure between the fully optimized PBE and HSE structures, compared to those optimized with fixed lattice constants, do not lead to significant deviations in the HSE+SOC band structures (Figure~\ref{fig:HDP_cell-relax} (c)).

Due to the relatively more compacted Ge 4$s$ states compared to the Sn 5$s$ and Pb 6$s$ states, Ge-doped HPs typically exhibit more pronounced stereochemical activity\cite{Swift/Lyons:2023, Fabini/etal:2024, Li/etal/2024}. For example, in 2D HPs, this can weaken or even reverse the trend of the band gap narrowing with increasing layer thickness\cite{Li/etal/2024}. Here, we construct a $3\times 3\times 3$ supercell of CsSnI$_3$ and randomly replace one Sn atom with Ge to study the stereochemical effects of Ge as an impurity, using PBE and HSE functionals. For the case where only atomic positions are optimized, the lattice parameters are fixed at $a = b = c = 19.311$ \AA~and $\alpha = \beta = \gamma = 90.0^\circ$. When allowing for cell relaxation, similar to the unit cell calculations, the lattice parameters predicted by PBE and HSE both decrease. Specifically, they are $a = b = c = 18.925$ \AA, $\alpha = \beta = \gamma = 89.0^\circ$ for PBE, and $a = b = c = 18.847$ \AA, $\alpha = \beta = \gamma = 89.2^\circ$ for HSE. 

\begin{figure}[!htbp]
	\includegraphics[width=\textwidth]{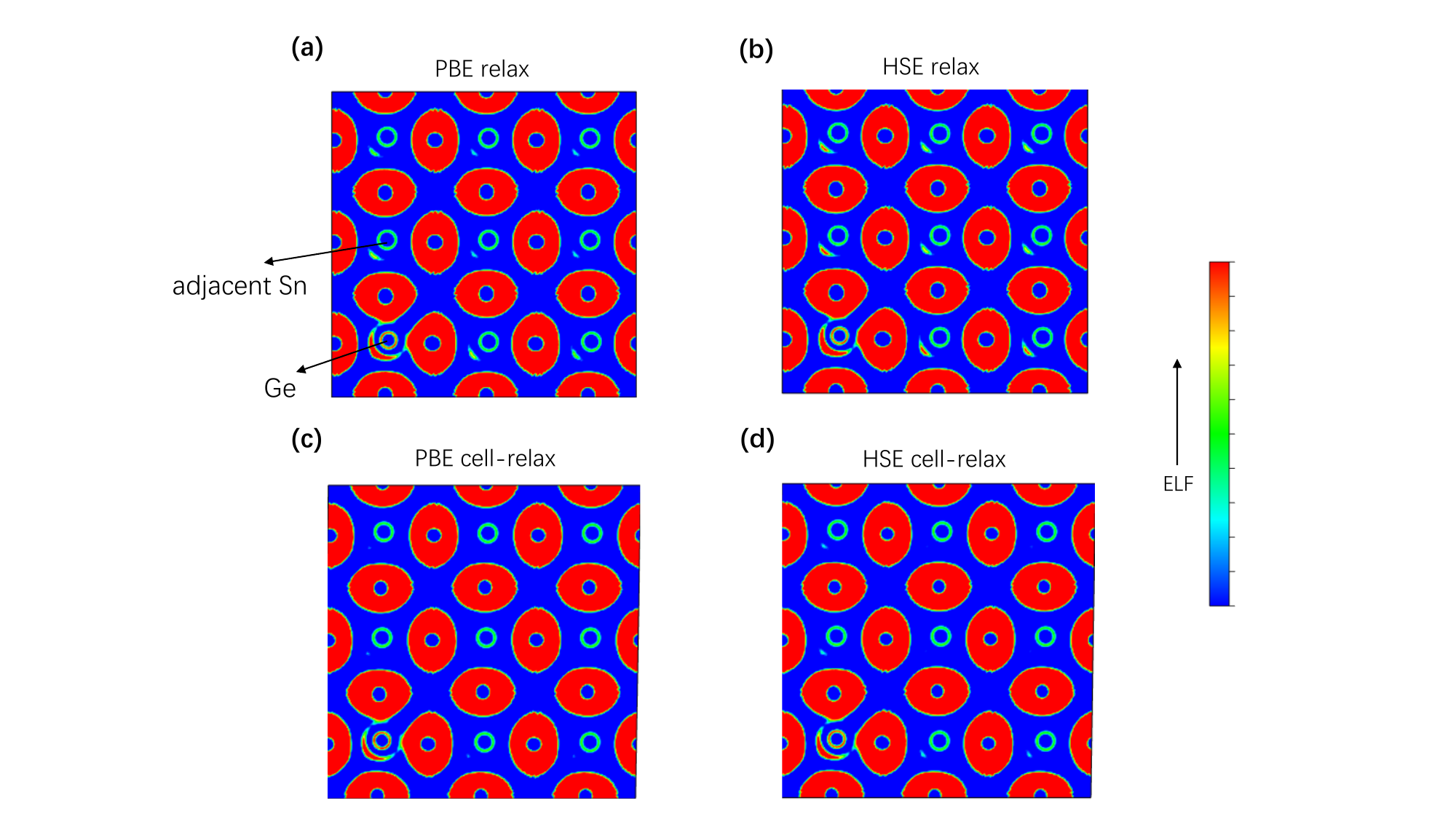}
	\centering
    \caption{(a)-(b) ELFs on the (001) plane for the relaxed PBE and HSE Ge-doped CsSnI$_3$ structures, with fixed lattice parameters, respectively. (c)-(d) ELFs on the (001) plane for the fully relaxed PBE and HSE Ge-doped CsSnI$_3$ structures, with both atomic positions and lattice parameters optimized, respectively.
	\label{fig:HDP_ELF}}
\end{figure}

According to the ELFs (Figure~\ref{fig:HDP_ELF}) and octahedral distortion parameters (Table~\ref{tab:HDP_333}), the octahedron centered on Ge exhibits more significant distortion compared to the one centered on the adjacent Sn. Although the lattice is compressed after optimization, the distortion does not weaken, which contrasts with the behavior observed in the unit cell case. This can be attributed to the fact that doping inherently leads to a local symmetry breaking, which prevents the distortion from diminishing.

\begin{table}[!htbp]
    \centering
    \caption{Comparison of octahedral distortion parameters for the atomic position-relaxed and fully relaxed Ge-doped CsSnI$_3$ supercell, optimized using PBE and HSE. Two types of off-centering are presented in the ``Center" column: Ge and adjacent Sn.}
    \label{tab:HDP_333}
    \begin{tabular}{c | c | c | c | c | c }
        \toprule 
            Method & Center & $d^0_\text{Sn-I}$(\AA) & $\sigma_1^2$ (deg$^2$) & $\sigma_2^2$ (deg$^2$) & $\lambda$ \\
        \midrule 
            \multirow{2}{*}{PBE relax} & Ge & 3.144 &	43.6 &	181.6 &	1.011 
\\ 
                    & adjacent Sn & 3.241 &	12.5 &	68.5 	& 1.005 
\\
        \midrule
            \multirow{2}{*}{HSE relax} & Ge & 3.151 &	45.6 &	190.1 & 1.015 
 \\
                    & adjacent Sn & 3.241 &	16.3 &	81.0&1.004 
 \\
        \midrule
            \multirow{2}{*}{PBE cell-relax} & Ge & 3.083 &	38.2 	&203.3 &	1.007 
\\
                    & adjacent Sn & 3.175 &	14.6 &	75.9 &	1.002 
\\
        \midrule
            \multirow{2}{*}{HSE cell-relax} & Ge & 3.065 &	44.8 &	242.2 &	1.040 
\\
                    & adjacent Sn & 3.154 &	21.5 &	116.5 &	1.003 
\\
        \bottomrule 
    \end{tabular}
\end{table}

\begin{figure}[!htbp]
	\includegraphics[width=\textwidth, clip, trim=3cm 3cm 3cm 3cm]{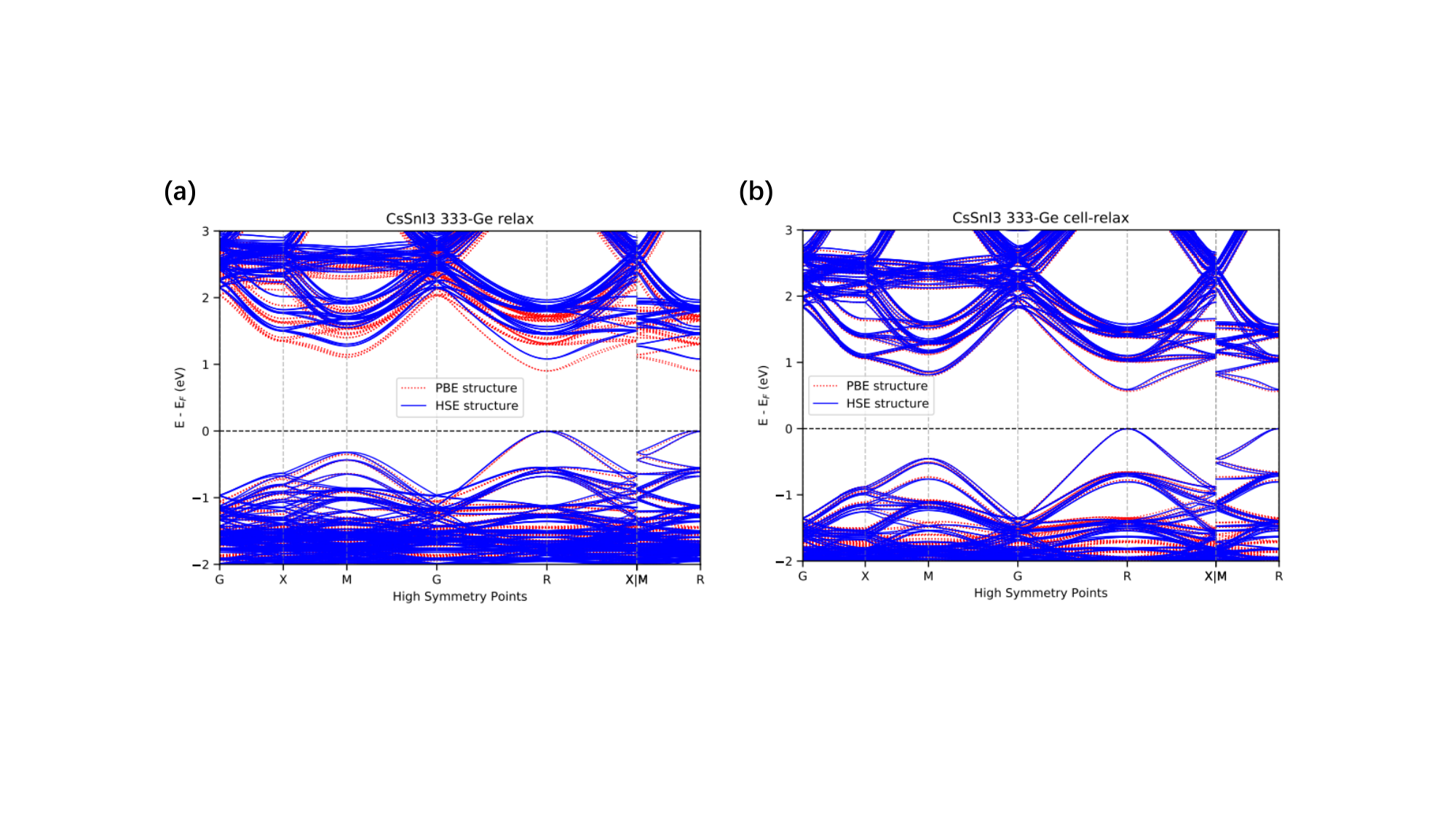}
	\centering
    \caption{(a)-(b) HSE+SOC band structure for the atomic position-relaxed
    and fully relaxed Ge-doped CsSnI$_3$ supercell. 
	\label{fig:Dope_cell-relax}}
\end{figure}

Figure~\ref{fig:Dope_cell-relax} (a) depicts the HSE+SOC band structures of the Ge-doped supercell with atomic position optimization based on PBE and HSE. Under a certain strain, PBE underestimates the stereochemical effects of the Ge and Sn $s^2$ lone pairs, leading to a reduction in the band gap. For the fully optimized case, a comparison of Figure~\ref{fig:HDP_ELF} (c) and (d) shows that both PBE and HSE can effectively describe the local symmetry breaking induced by the impurity. However, due to spatial constraints, no significant off-centering of Sn is observed, and as a result, the HSE+SOC band structures corresponding to the PBE and HSE optimized structures show little difference.

\section{Summary\label{sec:summary}}
In summary, we have developed an efficient real-space algorithm for evaluating the atomic forces and stress tensors derived from the HFX energy within the NAO basis-set framework. By exploiting 
the sparsity of the expansion coefficients under the LRI approximation and the density matrix,
the algorithm features a linear scaling of the computational cost with respect to the system size. 
We implemented the algorithm within the NAO-based ABACUS code package in a massively parallelized way via a hybrid MPI/OpenMP scheme. Scaling tests with respect to system size and computational resources were performed, and linear scaling behavior of the computational cost with respect to system size, as well as the excellent parallel efficiency, were observed. Our implementation enables efficient structure relaxation for large-scale systems at the level of hybrid functional.
Finally, we calculate the HSE+SOC band structures of the halide perovskite material CsSnI$_3$
and its Ge-doped counterpart with geometries relaxed using the PBE and HSE functionals, and found that under strain the band gap is appreciably larger at the HSE atomic structures, while this
difference diminishes with fully relaxed cells.
\begin{suppinfo}

The following data are available in the Supporting Information.
\begin{itemize}
  \item The test results on the dependence of the force errors and computation times on the 
  threshold parameters for 21 crystalline materials. 
\end{itemize}

\end{suppinfo}

\begin{acknowledgement}
  We acknowledge the funding support by National Natural Science Foundation of China
  (Grant Nos 12134012, 12374067, 12188101 and 12204332),
  Guangdong Basic and Applied Basic Research Foundation (Project Numbers 2021A1515110603).
  This work was also supported by the Strategic Priority Research Program of Chinese Academy of Sciences under Grant No. XDB0500201
and by the National Key Research and Development Program of China (Grant Nos. 2022YFA1403800 and 2023YFA1507004).
The numerical calculations in this study were partly carried out on the TianGong Supercomputer and ORISE Supercomputer.  
\end{acknowledgement}
\bibliography{CommonBib}
\end{document}